\documentclass{emulateapj}
\usepackage{graphics, ulem}
\usepackage{psfig,epsf}
\begin{document}
\baselineskip=12pt


\title{DUSTY INFRARED GALAXIES: Sources of the Cosmic Infrared Background}
\markboth{\sc Lagache, Puget  \& Dole}{\sc Sources of the Cosmic Infrared Background}

\author{Guilaine Lagache, Jean-Loup Puget, Herv\'e Dole}
\affil{Institut d'Astrophysique Spatiale, B\^at 121, Universit\'e
de Paris Sud 11, 91405 Orsay Cedex, France; e-mail: guilaine.lagache@ias.u-psud.fr, 
puget@ias.u-psud.fr, herve.dole@ias.u-psud.fr}

\shorttitle{Dusty IR Galaxies: Sources of the Cosmic IR Background --
ARAA 2005}
\shortauthors{Lagache, Puget, Dole -- ARAA 2005}

\slugcomment{To Appear in Annual Reviews of Astronomy \& Astrophysics, 2005, volume 43}

\keywords{cosmology -- evolution -- luminosity function -- starburst -- star formation}

\begin{abstract}
The discovery of the Cosmic Infrared Background (CIB) in 1996,
together with recent cosmological surveys from the mid-infrared to the
millimeter have revolutionized our view of star formation at high
redshifts.  It has become clear, in the last decade, that a population
of galaxies that radiate most of their power in the far-infrared (the
so-called ``infrared galaxies'') contributes an important part of the
whole galaxy build-up in the Universe. Since 1996, detailed (and often
painful) investigations of the high-redshift infrared galaxies have
resulted in the spectacular progress covered in this review. We
outline the nature of the sources of the CIB including their
star-formation rate, stellar and total mass, morphology, metallicity
and clustering properties. We discuss their contribution to the
stellar content of the Universe and their origin in the framework of
the hierarchical growth of structures.  We finally discuss open
questions for a scenario of their evolution up to the present-day
galaxies.
\end{abstract}

\section{INTRODUCTION}

The Cosmic Infrared Background (CIB) can be defined as the part of the
present radiation content of the Universe that is made essentially of
the long wavelength output from all sources throughout the history of
the Universe. The radiation content in the microwave part of the
spectrum is dominated by the Cosmic Microwave Background (CMB)
produced in the hot and dense phases of the universe. It dominates for
frequencies below 800 GHz. Nevertheless the very different spectra of
the CIB with respect to the CMB (both its purely Planckian part and
its Compton distortion expected to be the dominant one at these
frequencies) allow them to be separated very efficiently down to
frequencies close to the peak of the CMB (150 GHz). Furthermore in
this frequency regime the CIB dominates the galactic emission in the
lowest cirrus regions by a factor $\simeq$4.  The cosmic background
due to sources (CMB excluded) presents two maxima: one in the optical,
one in the far-infrared, with roughly equal brightness (in $\nu
I_{\nu}$) and with a minimum around 5~$\mu$m. This minimum is created
by the decrease of brightness of the stellar component with wavelength
combined with the rising brightness of the dust, very small grains,
and of the Active Galactic Nuclei (AGN) non thermal emission in the
thermal infrared.  The CIB is defined as the cosmic background at
wavelengths longward of this minimum.  An understanding of the nature
and redshift distribution of the sources of the CIB, although
relatively new, is an integral part of the understanding of the
formation and evolution of galaxies.

The standard hierarchical model of structure formation has received
strong observational support from observations of the large-scale
distribution of galaxies, clusters, intergalactic clouds, combined
with CMB anisotropies that constrain the initial large scale power
spectrum within the concordance cosmological model framework.
Scenarios for galaxy formation and evolution can be confronted with
the very quickly rising set of observations of extragalactic sources
at higher and higher redshifts.  Nevertheless many critical questions
remain open on the cooling of collapsed structures, angular momentum
of galaxies, star formation efficiency, Initial Mass Function (IMF) of
the stars formed, role of feedback mechanisms, the physics and role of
merging and accretion in the construction of galaxies.

As ultraluminous infrared galaxies were found to be often associated
with mergers or interacting galaxies, it can be expected that the
sources of the CIB carry critical information about the history of
merging (e.g., Sanders \& Mirabel 1996; Genzel \& Cesarsky 2000).
Because about half of the energy from extragalactic sources is in the
CIB, the determination of the source of this energy (starbursts or
massive black hole accretion in dust enshrouded AGNs) should shed some
light on how galaxies evolve.

Since the discovery of the CIB in the COBE data (e.g., Hauser \& Dwek
2001), identifying the sources of the CIB, their redshift
distribution, and nature progressed at increasing speed especially
through multiwavelength analysis.  This review attempts to give the
broad-band observational picture for the identification of the sources
of the CIB.  The detailed analysis of individual infrared galaxies is
outside the scope of this review.  We then discuss implications, both
well-established ones and tentative ones, as well as directions for
future work. The spectroscopy aspects are covered in the review by
Solomon \& Vanden Bout (2005, in this volume).  This is at present a
fascinating but moving target! The coming decade will be a very rich
one. New, very powerful long-wavelengths observation tools will bring
many striking new results like the {\it Spitzer} observatory or the
being built experiments like {\it Herschel} and ALMA. Throughout this
review, the cosmology is fixed to $\rm \Omega_{\Lambda}$=0.7, $\rm
\Omega_{m}$=0.3, H$_0$=100 h km s$^{-1}$ Mpc$^{-1}$ with h=0.65.

\section{DUST IN THE LOCAL UNIVERSE}

A fraction of the stellar radiation produced in galaxies is absorbed
by dust and re-radiated from mid-infrared to millimeter wavelengths.
Understanding dust properties and the associated physics of the absorption
and emission are thus essential. These determine the Spectral Energy
Distribution (SED) of the galaxies. 

\begin{figure*}[!ht]
\begin{center}
\includegraphics[width=1.0\linewidth]{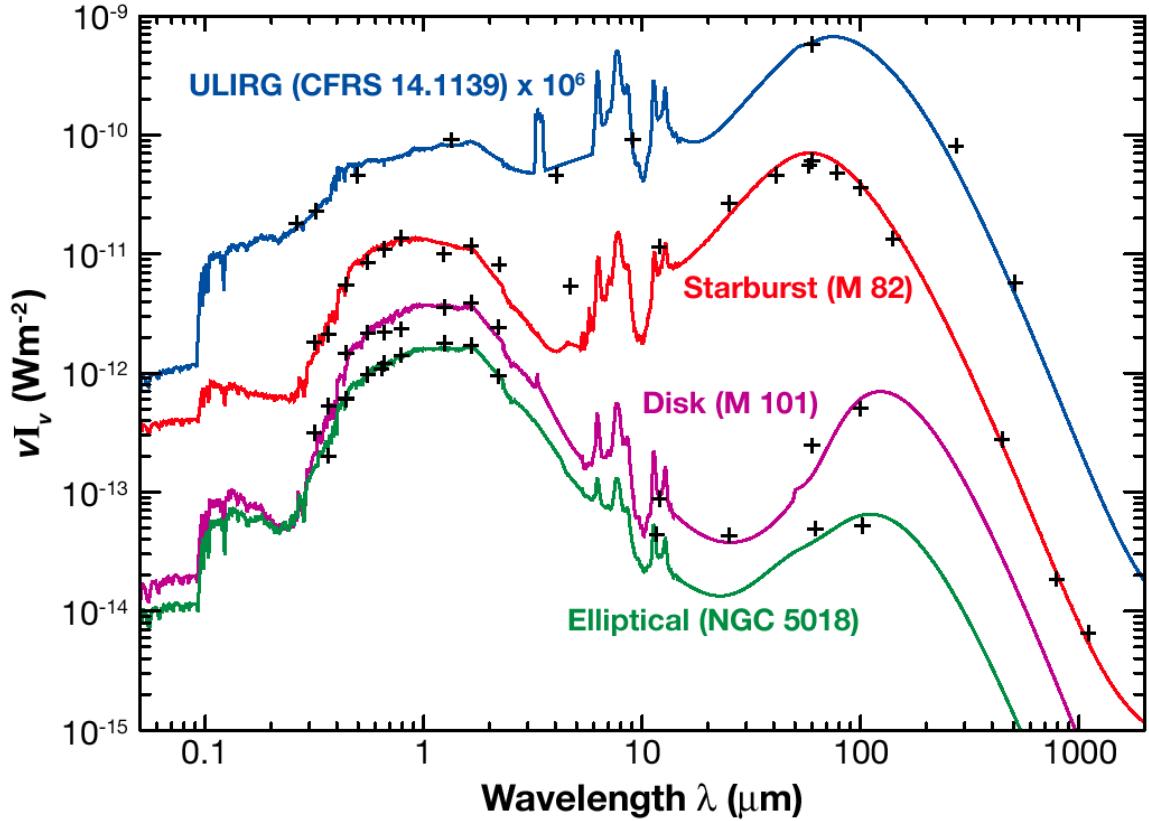}
\end{center}
\caption{Spectral energy distributions of galaxies from UV to the millimeter. The ULIRG 
is observed at redshift z=0.66 and is represented here
in the rest-frame (from Galliano 2004).
}
\label{fig:gal_spectra}
\end{figure*}

\subsection{Dust Particles}

Small dust particles with sizes ranging from a nanometer to a fraction
of micrometer are ubiquitous in the interstellar medium.  They result
from natural condensation in cool stellar atmospheres, supernovae, and
the interstellar medium of the heavy elements produced by the
nucleosynthesis in stars and released to the diffuse medium by late
type stars and supernovae explosions. Interstellar grain models have
been improved for 30 years in order to fit all observational
constraints: elemental abundances of the heavy elements, UV, visible
and infrared absorption and scattering properties, infrared emission,
polarization properties of the absorbed and emitted light.  The models
include a mixture of amorphous silicate grains and carbonaceous
grains, each with a wide size distribution ranging from molecules
containing tens of atoms to large grains $\ge$ 0.1 $\rm \mu$m in
diameter that can be coated with ices in dense clouds and/or organic
residues (e.g., D\'esert et al. 1990; Li \& Draine 2001). It is now
widely accepted that the smallest carbonaceous grains are Polycyclic
Aromatic Hydrocarbons (PAHs) that emit a substantial fraction of the
energy in a set of features between 3 and 17~$\rm \mu m$ (3.3, 6.2,
7.7, 8.6, 11.3, 12.7, 16.3, 17 $\rm \mu$m for the main ones) that used
to be known as the UIB for Unidentified Infrared Bands.  These
features result from C-C and C-H stretching/bending vibrational bands
excited by the absorption of a single UV or optical photon and are a
good tracer of normal and moderately active star formation activity in
spiral and irregular galaxies (e.g., Helou et al. 2000; Peeters et
al. 2004). For radii a$\ge$50 ${\rm \AA}$, the carbonaceous grains are
often assumed to have graphitic properties. The so-called very small
grains of the interstellar medium are small enough to have very low
heat capacity, so their temperature are significantly affected by
single-photon absorption. In the diffuse ISM of our Galaxy, they
dominate the infrared emission for wavelengths smaller than about 80
$\rm \mu$m.  At longer wavelengths, the infrared spectrum is dominated
by the emission of the larger grains at their equilibrium
temperatures. Considering the energy density of the radiation in a
galactic disc like ours, the temperature of the larger grains is
rather low; 15 to 25 K. For these grains, the far-infrared emissivity
decreases roughly as the square of the wavelength. This in turn makes
the temperature dependence on the radiation energy density $u$ very
weak ($ T \simeq u^{1/6}$). For a galaxy like the Milky Way, the
infrared part of the SED peaks at 170 $\rm \mu m$ whereas for an Ultra
Luminous Infrared Galaxy (ULIRG) it peaks at about 60 $\rm \mu m$: a
factor $3$ in temperature for a factor $10^3$ in energy density or
luminosity. At long wavelengths in the submillimeter and millimeter,
the intensity should increase like $I_\nu \simeq \nu^{4}$.

\subsection{Extinction}

In our Galaxy, the extinction curve of the diffuse ISM has been known
for a few decades. The average optical depth perpendicular to the disk
of our Galaxy in the solar vicinity is small (A$_V$ $\simeq 0.2$) and
typical of spiral galaxies. The average optical depth increases to a
few in large molecular cloud complexes. It can become very large in
galactic nuclei. Finally it should also be remembered that the optical
depth in the UV is typically 10 times larger than that in optical
wavelengths. The conversion of star light into infrared radiation will
thus depend strongly on the location of the stars and their spectral
types.

In external galaxies, modeling the extinction is very hard because it
strongly depends on the geometric distribution of the ISM and of the
chemical abundances. Simple models have been used to take this into
account to first order.  Galaxies can be modeled as an oblate
ellipsoid where absorbers (dust) and sources (stars) are homogeneously
mixed; the dust absorption can be computed in a ``screen'' or
``sandwich'' geometry (dust layers in front of the stars or sandwiched
between two star layers). As a consequence, the reddening curve
average over a whole galaxy appears to vary within a class of objects
and between the different classes, from normal star-forming galaxies
to highly concentrated starburst.  It is thus very difficult to derive
the total dust optical depth (e.g., Calzetti et al. 1994). In the
local Universe, the average extinction per galaxy is quite low. About
one third of the bolometric flux is emitted in the far-infrared, and
this is typical of our Galaxy. In more actively star-forming galaxies,
up to 70\% of the bolometric flux is emitted in the far-infrared. In
some of these, the starburst activity is mostly in the disk (like in
M51). For a given total luminosity, the radiation energy density is
lower than in the case of a starburst concentrated in a small volume
in the nucleus. In this case, the dust will be hotter due to the
larger energy density and the conversion of stellar light to infrared
will be more efficient. Some ULIRGs emit more than 95\% of their
energy in the far-infrared (e.g., Arp 220). Such galaxies are very
compact, dusty starbursts where dust optical depths are very large. In
such galaxies, fine structure and recombination line ratios imply an
equivalent ``screen'' dust extinction between A$_{\rm v}$$\sim$5 and
50. The result is that the SED is significantly distorted in the
opposite way from the higher dust temperature (less mid-infrared
emission).  In the following, we will refer to ``infrared galaxies''
and to ``optical galaxies'' to designate galaxies in which the
infrared emission, respectively optical emission dominates. Different
typical spectra of galaxies are shown in Figure \ref{fig:gal_spectra}
from the UV to the millimeter.  We clearly see the variation of the
optical to infrared energy ratio as starburst activity increases.

\subsection{Local Infrared Galaxies}

A few very luminous infrared galaxies were observed in the seventies
(Rieke \& Lebofsky 1979). Then IRAS satellite, launched in 1983 gave
for the first time a proper census of the infrared emission of
galaxies at low redshift.  The Luminosity Function (LF) at 60 and 100
$\rm \mu m$ is dominated by $L_\star$ spiral galaxies as could be
expected -- the reradiated stellar luminosity absorbed by dust.  In
addition, a high-luminosity tail of luminous galaxies was found (e.g.,
Sanders \& Mirabel 1996). This high-luminosity tail can be
approximated by a power-law, $\Phi (L) \propto L_{\rm IR}^{2.35}$,
which gives a space density for the most luminous infrared sources
well in excess of predictions based on the optical LF. These sources
comprise the Luminous Infrared Galaxies, LIRGs, and the ULIRGs with
luminosities 11$<$log(L$_{\rm IR}$/L$_{\odot}$)$<$12 and log(L$_{\rm
IR}$/L$_{\odot}$)$>$12, respectively. These galaxies are often
associated with interacting or merging, gas-rich disks.  The fraction
of strongly interacting/merger systems increases from $\sim$10\% at
log(L$_{\rm IR}$/L$_{\odot}$)=10.5-11 to $\sim$100\% at log(L$_{\rm
IR}$/L$_{\odot}$)$>$12.  LIRGs are the site of intense starburst
activity (about 10-100 M$_{\odot}$ year$^{-1}$) induced by the
interaction and/or strong spiral structure.  The ULIRG phase occurs
near the end of the merging process when the two disks overlap. Such
galaxies may be the precursors of Quasi Stellar Objects (QSOs; Sanders
et al. 1988a, 1988b; Veilleux et al. 1995; Lutz et al. 1999). These
objects have been the subject of intense debate concerning the nature
of the dominant source of emission: starburst versus dust-enshrouded
AGN (e.g., Filipenko 1992; Sanders \& Mirabel 1996; Joseph 1999).
Indeed, spectra show evidence of extremely large optical depth
(heavily reddened continuum and large Balmer decrement) but also
exhibit AGN-like high excitation fine-structure lines. We had to wait
for ISO to clearly determine the power sources of ULIRGs. The
difference between the mid-infrared spectra of starburst and AGNs is
striking. Starburst are often characterized by strong, low-excitation,
fine-structure lines, prominent PAH features and a weak $\lambda
\ge$10 $\rm \mu$m continuum whereas AGNs display a highly excited
emission line spectrum with weak or no PAH features, plus a strong
mid-infrared continuum.  It has been thus possible to build
mid-infrared diagnostic diagrams (e.g., Genzel et al. 1998; Laurent et
al. 2000) that clearly separates starburst-dominated galaxies from
AGN-dominated galaxies. These diagrams demonstrate that ULIRGs appear
to be composite objects, but star formation dominates in most
objects. That is on average, $\ge$70\% of the reradiated energy comes
from starbursts and $\le$ 30\% comes from AGNs (Genzel et al. 1998;
Lutz et al. 1998). However the fraction of AGN-powered objects
increases with luminosity. About 15$\%$ of ULIRGs at luminosities
below 2$\times$10$^{12}$ L$_{\odot}$ are mostly AGN powered, but this
fraction increases to about half at higher luminosity.

All these well-studied LIRGs and ULIRGs are at low
redshift. They do not dominate the energy production locally.
As an example, the total infrared luminosity from these
galaxies in the IRAS Bright Galaxy Sample accounts for only 
$\sim$6\% of the infrared emission in the local Universe
(Soifer \& Neugeubauer, 1991). As we will see, the situation
changes dramatically at higher redshift where
these galaxies fully dominate the infrared energy output.

\begin{figure*}[!ht]
\begin{center}
\includegraphics[width=1.0\linewidth]{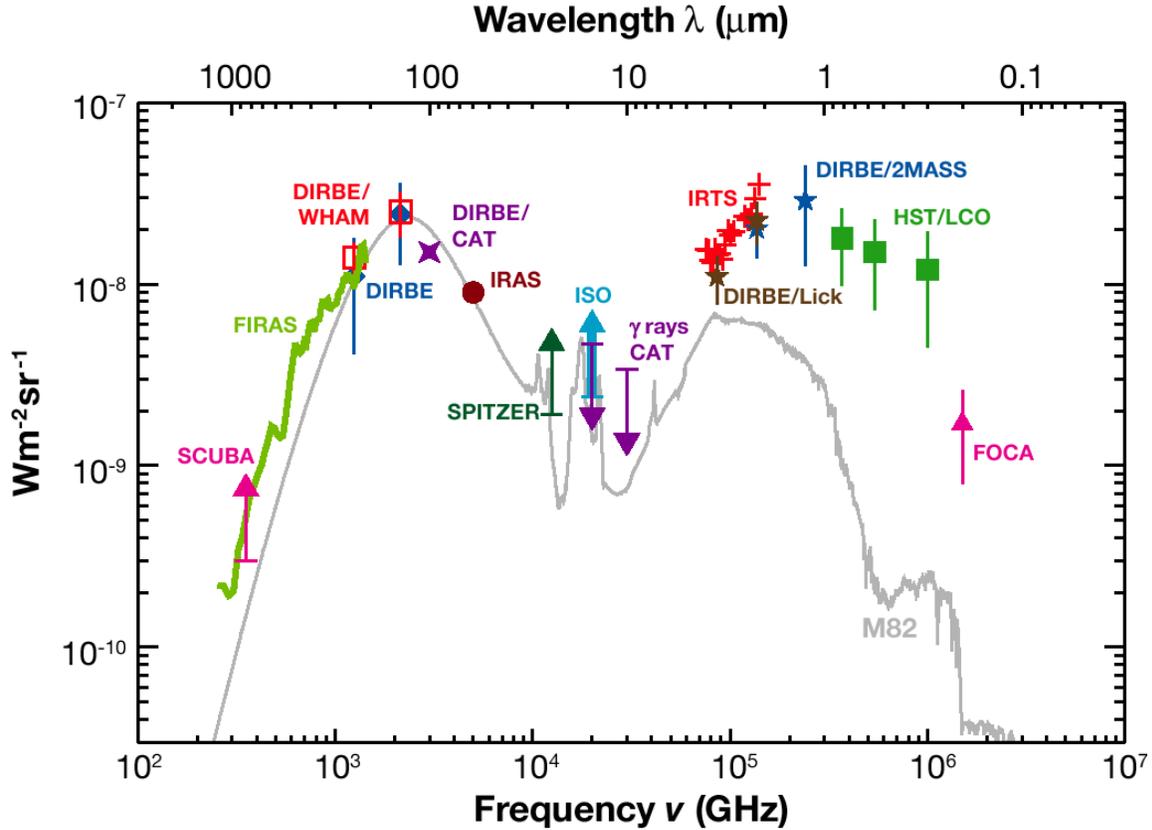}
\end{center}
\caption{The extragalactic background over three decades in frequency from the
near UV to millimeter wavelengths. Only strongly constraining
measurements have been reported. We show for comparison in grey an SED
of M82 (Chanial, 2003)-- a starburst galaxy at L=3$\times$ 10$^{10}$
L$_{\odot}$ -- normalized to the peak of the CIB at 140~$\mu$m. 
References for data points are given in Table \ref{table:CIB}.
\label{fig:CIB_tot}}
\end{figure*}

\section{THE COSMIC INFRARED BACKGROUND}

The CIB is the infrared part of the extragalactic background, the
radiation content of the Universe today produced by galaxies at all
redshifts and seen as an isotropic extragalactic background
radiation. Patridge \& Peebles (1967) predicted that observations of
such a background could give powerful constraints on the cosmological
evolution.

\subsection{General Observations and Direct Cosmological Implications}
The detection of the infrared part of the extragalactic background
(the CIB for Cosmic Infrared Background) was the major objective of
the DIRBE experiment aboard COBE. In fact, the CIB was first detected
at long wavelengths by using the FIRAS spectrometer: $\rm \lambda >$
200 $\rm \mu$m (Puget et al. 1996). The CIB has subsequently been
detected by DIRBE at 2.4, 3.5, 100, 140, 240 $\rm \mu$m (see Hauser \&
Dwek 2001 and Kashlinsky 2005 for two reviews). The extragalactic
background at 2.4 and 3.5~$\mu$m is significantly larger than that
predicted by the integrated galaxy counts and their extrapolation.
Similarly, the extragalactic background in the optical has been
finally evaluated in combining several methods by Bernstein et
al. (2002) and found to be larger than the value given by the
integrated fluxes of galaxies by a factor larger than 2.  In the
mid-infrared, the interplanetary zodiacal dust emission is so strong
that only upper limits were obtained by DIRBE. The combination of
number counts by ISO/ISOCAM at 15 $\rm \mu$m (see Elbaz \& Cesarsky
2003) and by {\it Spitzer} at 24~$\rm \mu$m (e.g., Papovich et
al. 2004) giving lower limits, with the observations of TeV gamma ray
emission from distant AGNs (e.g., Renault et al.  2001; Dwek \&
Krennrich 2005), gives a good measurement of the background at these
wavelengths. The full cosmic background spectrum is shown in Figure
\ref{fig:CIB_tot}. Only most recent and strongly constraining
measurements have been plotted for clarity. \\

Figure \ref{fig:CIB_tot} clearly shows that the optical and infrared
cosmic backgrounds are well separated. The first surprising result is
that the power in the infrared is comparable to the power in the
optical. In contrast, we know that locally, the infrared output of
galaxies is only one third of the optical output. This implies that
infrared galaxies grow more luminous with increasing $z$ faster than
do optical galaxies.  A second important property to note is that the
slope of the long wavelength part of the CIB, $I_\nu \propto
\nu^{1.4}$ (Gispert et al. 2000), is much less steep than the long
wavelengths spectrum of galaxies (as illustrated in Figure
\ref{fig:CIB_tot} with the M82 SED).  This implies that the millimeter
CIB is not due to the millimeter emission of the galaxies that account
for the peak of the CIB ($\simeq 150 \mu$m). The implications in terms
of energy output have been drawn by, e.g. Gispert et al. (2000). The
infrared production rate per comoving unit volume $(a)$ evolves faster
between redshift zero and 1 than the optical one and $(b)$ has to stay
roughly constant at higher redshifts up to redshift 3 at least.

\subsection{\label{surveys} The Status of Deep Surveys: Resolved Fraction of the $>$10 $\rm \mu$m CIB}

Many surveys from the mid-infrared to the millimeter have aimed to
resolve the CIB into discrete sources.
From short to long wavelengths the significant surveys are the following:
\begin{itemize}

\item ISOCAM 15 $\rm \mu$m: Three kinds of surveys have been done. The shallowest
is the ELAIS survey (European Large-Area ISO Survey, Oliver et
al. 2000). The deepest is the survey in the HDF-N (Aussel et al. 1999)
as well as the surveys in the direction of the galaxy clusters
(e.g. Metcalfe et al. 2003). Altogether, about 1000 galaxies were
detected. Above a sensitivity limit of 50 $\rm \mu$Jy, they produce a
15 $\rm \mu$m extragalactic background light of (2.4 $\pm$ 0.5) nW
m$^{-2}$ sr$^{-1}$ (Elbaz et al. 2002). This accounts for about 80\%
of the CIB at 15 $\rm \mu$m based on the simplest extrapolation of the
counts.

\begin{deluxetable*}{lccc}
\tablewidth{0pt}
\tablecaption{Extragalactic background references for Figure \ref{fig:CIB_tot}
\label{table:CIB}}
\tablehead{
\colhead{{\bf Wavelength ($\rm \mu$m)}} &
\colhead{{\bf Experiment}} &
\colhead{{\bf Measurement}} &
\colhead{{\bf Reference}} \\
}
\startdata
0.2 & FOCA & Number counts \& model & Armand et al. 1994 \\
0.30, 0.56, 0.81 & HST/Las  & Diffuse emission & Bernstein et al. 2002 \\
& Campanas Obs. & &  Mattila 2003 \\
2.2$< \lambda < $4 & IRTS & Diffuse emission & Matsumoto et al. 2005 \\
2.2, 3.3 & DIRBE/Lick & Diffuse emission & Gorjian et al. 2000 \\
1.25, 2.2 & DIRBE/2MASS & Diffuse emission & Wright 2001 \\
& & & Cambr\'esy et al. 2001 \\
10, 15 & CAT & $\gamma$-rays & Renault et al. 2001 \\
15 & ISO/ISOCAM & Number counts & Elbaz et al. 1999 \\
24 & {\it Spitzer}/MIPS & Number counts & Papovich et al. 2004 \\
60 & IRAS & Power spectrum & Miville-Desch\^enes et \\
& & & al. 2001 \\
100 & DIRBE & Diffuse emission & Renault et al. 2001 \\
140, 240 & DIRBE/WHAM & Diffuse emission & Lagache et al. 2000\\
140, 240 & DIRBE & Diffuse emission & Hauser et al. 1998\\
850 & SCUBA & Number counts & Smail et al. 2002\\
200$< \lambda < $1200 & FIRAS & Diffuse emission & Lagache et al. 2000\\
\enddata
\end{deluxetable*}

\item {\it Spitzer} 24 ~$\rm \mu$m: {\it Spitzer} surveys are ongoing, but 
more than 10$^5$ sources have already been detected at 24~$\rm
\mu$m. Integrating the first number counts down to 60 $\rm \mu$Jy, a
lower limit to the CIB at 24~$\rm \mu$m of (1.9$\pm$0.6) nW m$^{-2}$
sr$^{-1}$ is derived (Papovich et al. 2004). This accounts for about
70\% of the CIB at 24~$\rm \mu$m, based on a simple extrapolation of
the counts.

\begin{figure*}[!ht]
\begin{center}
\includegraphics[width=1.0\linewidth]{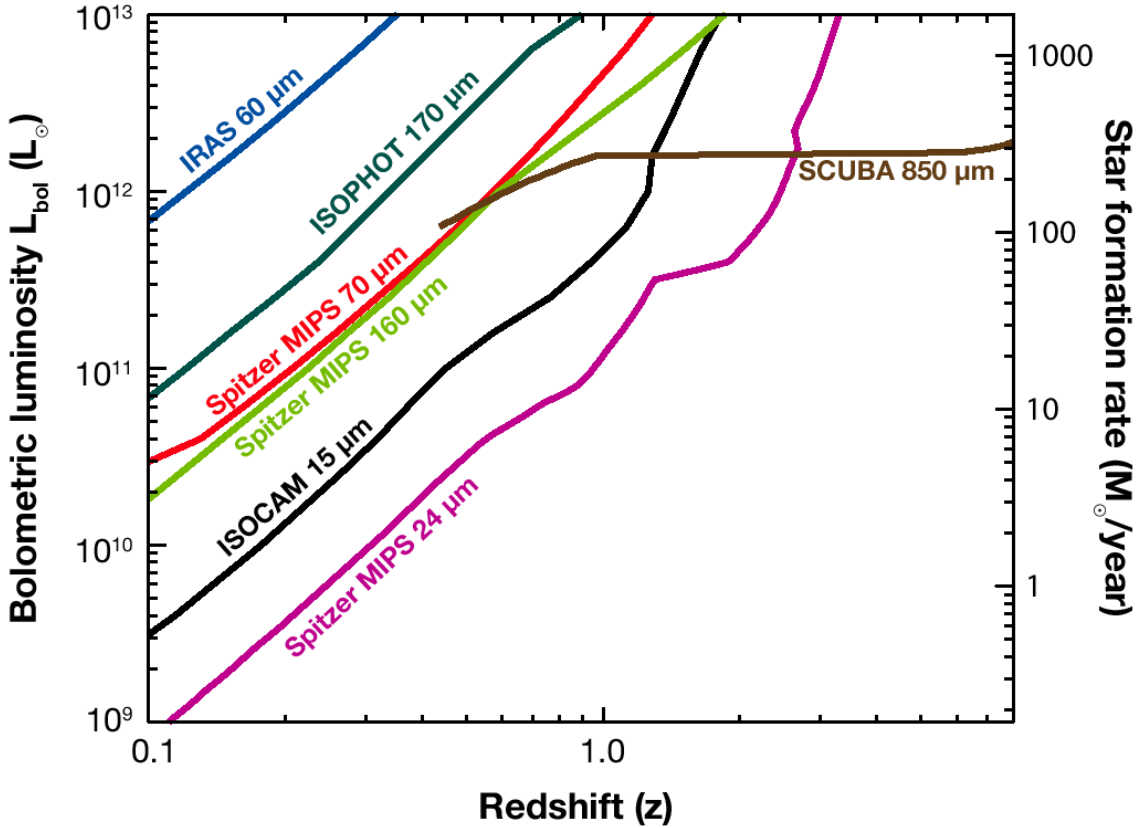}
\end{center}
\caption{Sensitivity to the bolometric luminosity and star-formation
rate, assuming star forming galaxies of various infrared and
submillimeter experiments. Detections of at least 10 sources in the
surveys can be expected in the areas above the {\it curves}.
We assumed the scenario of a typical deep survey (when available).
IRAS 60~$\mu$m ($S_{\nu} > 1$~Jy, all sky);
ISOCAM 15~$\mu$m ($S_{\nu} > 250 \mu$Jy, 2 Sq. Deg.);
ISOPHOT 170~$\mu$m ($S_{\nu} > 180 $~mJy, 5 Sq. Deg.);
{\it Spitzer}/MIPS 24~$\mu$m ($S_{\nu} > 80 \mu$Jy, 5 Sq. Deg.);
{\it Spitzer}/MIPS 70~$\mu$m ($S_{\nu} > 25 $~mJy, 5 Sq. Deg.);
{\it Spitzer}/MIPS 160~$\mu$m ($S_{\nu} > 50 $~mJy, 5 Sq. Deg.);
{\it SCUBA} 850~$\mu$m ($S_{\nu} > 1 $~mJy, 1 Sq. Deg.).
This plot makes use of the Lagache et al. (2004) model (see the Appendix).
}
\label{fig:plot_lbol_vs_z4araa}
\end{figure*}

\item {\it Spitzer} 70 $\rm \mu$m: Counts have been derived down to 15 mJy.
Integrating these counts corresponds 0.95 nW m$^{-2}$ sr$^{-1}$ which
explains $\sim$20\% of the CIB at 70~$\rm \mu$m as derived from the
Lagache et al. (2004) model (Dole et al. 2004a).

\item ISOPHOT 90 $\rm \mu$m: The most relevant data comes from
the ELAIS survey that covers about 12 square degrees at 90 $\rm \mu$m. 
Counts have been obtained down to 95 mJy (H\'eraudeau et al. 2004)
resolving less than 5$\%$ of the CIB as derived from DIRBE by Renault et al. 
(2001).

\item {\it Spitzer} 160 $\rm \mu$m: First counts are derived down to 50 mJy.
The integral of these counts corresponds to 1.4 nW m$^{-2}$ sr$^{-1}$,
which explain about 7\% of the CIB at 160 $\rm \mu$m (Dole et
al. 2004a).

\item ISOPHOT 170 $\rm \mu$m: Two main surveys have been conducted: the FIRBACK
(Lagache \& Dole 2001) and the Lockman hole (Kawara et al. 2004)
surveys that covering about 5 square degrees. Counts down to 135 mJy
contribute to less than 5\% of the CIB (Dole et al. 2001).

\item SCUBA 450 $\rm \mu$m: Deep surveys at 450 $\rm \mu$m are very hard to conduct
from the ground. A few galaxies are detected between 10 and 50 mJy in
deep surveys. Number counts down to 10 mJy give a lower limit on the
CIB of about (0.7 $\pm$ 0.4) nW m$^{-2}$ sr$^{-1}$ (Smail et
al. 2002). This resolves about 15\% of the CIB at 450 $\rm \mu$m.

\begin{figure*}[!ht]
\begin{center}
\includegraphics[width=1.0\linewidth]{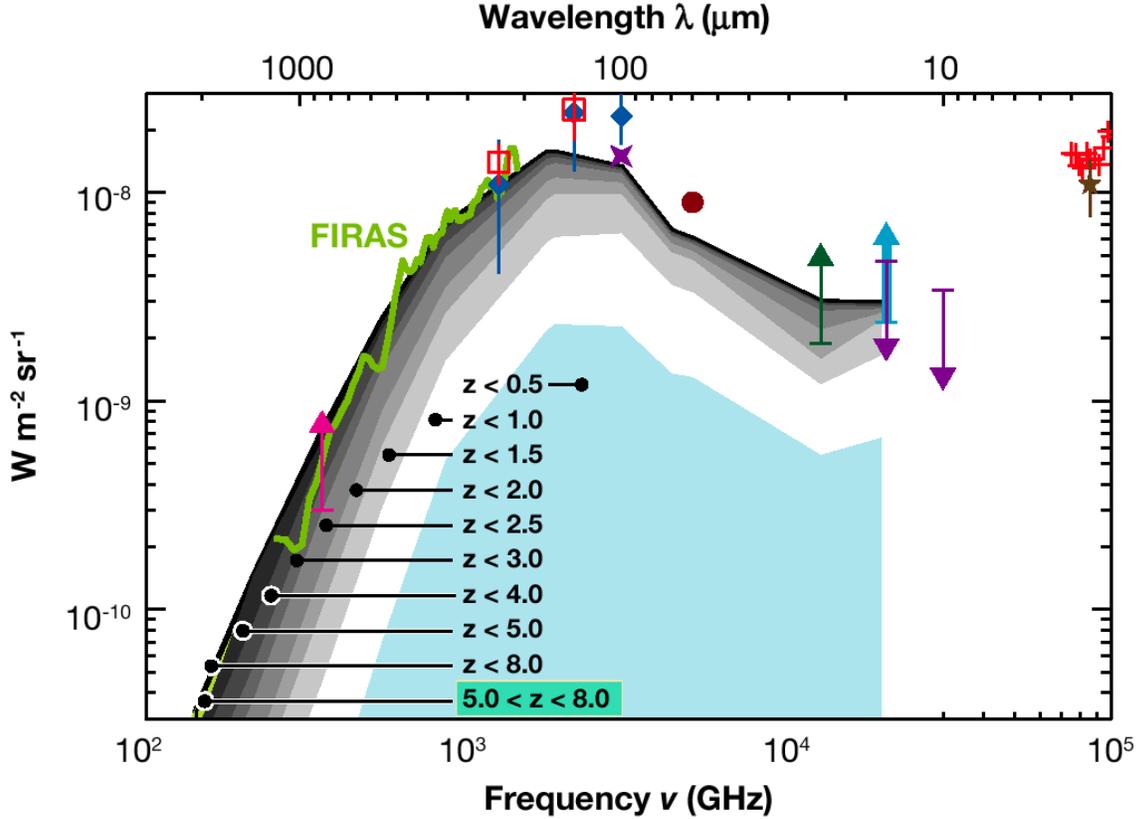}
\end{center}
\caption{Cumulative contribution to the CIB of galaxies at various
redshifts from 0.5 to 8, from the model of Lagache
et al. (2004). Measurements of the CIB are reported with the same
symbols as in Figure~\ref{fig:CIB_tot}.
}
\label{fig:CIB_fraction_spec}
\end{figure*}

\item SCUBA 850 $\rm \mu$m:  Over 500 arcmin$^2$ of blank sky has been surveyed 
by several groups using SCUBA. The observations range from an
extremely deep survey in the area of the HDF-N to a wider-field
shallower survey. Also about 40 arcmin$^2$ of lensed cluster fields
have been observed in which 17 sources have been detected.  Fluxes
range from about 0.5 to 8 mJy. The flux density in the resolved
submillimeter population down to 1 mJy is (0.3 $\pm$ 0.1) nW m$^{-2}$
sr$^{-1}$ (Smail et al. 2002). This account for 60$\%$ of the CIB at
850 $\rm \mu$m.  Note that in deep surveys, sources with
S$_{850}$$\ge$ 3 mJy contribute to $\sim$30\% of the CIB.

\item MAMBO 1200 $\rm \mu$m: Deep blank sky areas surveyed by MAMBO
cover about 500 arcmin$^2$. Number counts are given by Greve et
al. (2004). By integrating the counts from 2.25 to 5.75 mJy, the
resolved CIB is about 0.02 nW m$^{-2}$ sr$^{-1}$, or about 10\% of the
total CIB at 1200 $\rm \mu$m.

\end{itemize}

Figure \ref{fig:plot_lbol_vs_z4araa} shows the capabilities of the
different surveys to find distant LIRGs. {\it Spitzer} observations at
24~$\rm \mu$m are the most powerful tool to find LIRGs up to $z
\sim$2.2; ISOCAM was limited at $z \sim$1.2.  Distant ULIRGs are found
by deep and large surveys at 24 and 850~$\rm \mu$m.  Note that
capabilities have been computed using the model of Lagache et al.
(2004). This empirical model is based on only two populations of
galaxies; it aims only to model the redshift evolution of the average
population.  It reproduces all the observations from mid-infrared to
the millimeter (see Appendix).  Lewis et al. (2005) showed that a more
sophisticated, bivariate SED does not much change the average
properties although it does significantly change the dispersion.  The
Lagache et al. (2004) model is thus used in this paper as a tool to
discuss observations and predictions.

\begin{figure*}[!ht]
\begin{center}
\includegraphics[width=1.0\linewidth]{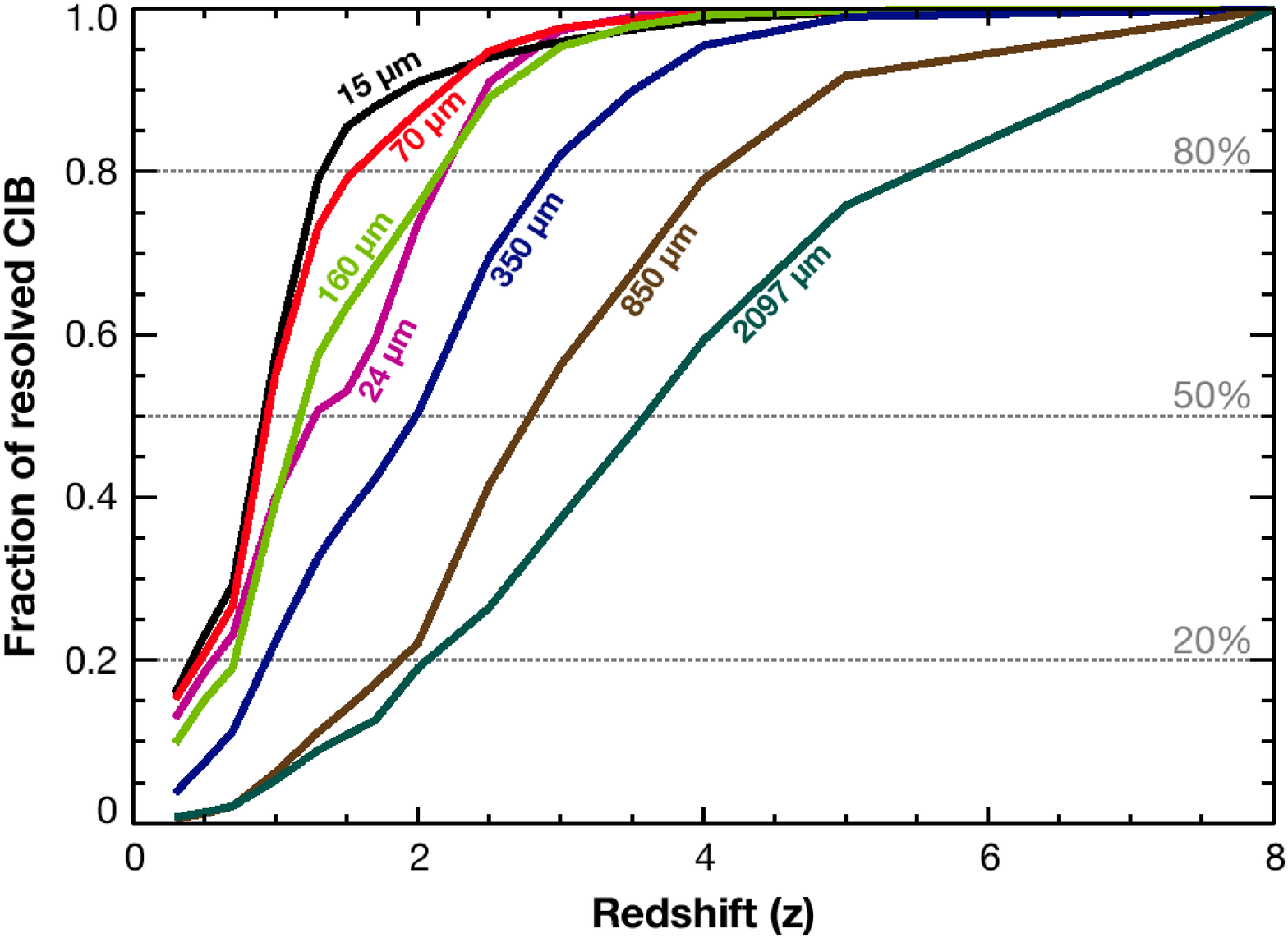}
\end{center}
\caption{Cumulative fraction of the CIB content
as a function of redshift for various wavelengths, from the model of
Lagache et al. (2004). 
}
\label{fig:cib_slice_z_cumul.color.ps}
\end{figure*}

\subsection{Redshift Contribution to the CIB}
From Figure \ref{fig:CIB_tot}, we see that contributions from galaxies
at various redshifts are needed to fill the CIB SED shape.  The bulk
of the CIB in energy, i.e., the peak at about 150 $\rm \mu$m, is not
resolved in individual sources but one dominant contribution at the
CIB peak can be inferred from the ISOCAM deep surveys. ISOCAM galaxies
with a median redshift of $\sim$0.7 resolve about 80\% of the CIB at
15~$\rm \mu$m. Elbaz et al. (2002) separate the 15 $\rm \mu$m galaxies
into different classes (ULIRGs, LIRGs, Starbursts, normal galaxies and
AGNs) and extrapolate the 15 $\rm \mu$m fluxes to 140~$\rm \mu$m using
template SEDs. A total brightness of (16$\pm$5) nW m$^{-2}$ sr$^{-1}$
is found, which makes up about two thirds of the CIB observed at 140
$\rm \mu$m by COBE/DIRBE. Hence, the galaxies detected by ISOCAM are
responsible for a large fraction of energy of the CIB. About one half
of the 140 $\rm \mu$m CIB is due to LIRGs and about one third to
ULIRGs. However, these ISOCAM galaxies make little contribution to the
CIB in the millimeter and submillimeter. There, the CIB must be
dominated by galaxies at rather high redshift for which the SED peak
has been shifted. The redshift contribution to the CIB is illustrated
in Figure \ref{fig:CIB_fraction_spec}.  We clearly see that the
submillimeter/millimeter CIB contains information on the total energy
output by the high-redshift galaxies ($z>$2). This is supported by the
redshift distribution of the SCUBA sources at 850 $\rm \mu$m with
S$_{850}$$\ge$3 mJy that make about 30\% of the CIB and have a median
redshift of 2.2 (Chapman et al. 2005).\\

\begin{deluxetable}{lccc}
\tablewidth{0pt}
\tablecaption{Redshift at which the Cosmic Infrared Background is
resolved at 20, 50 or 80\% (from the model
of Lagache et al. 2004). \label{tab:cib_resolution}}
\tablehead{
\colhead{Wavelength} &
\colhead{20\%} &
\colhead{50\%} &
\colhead{80\%} \\
}
\startdata
15~$\mu$m & 0.5 & 1.0 & 1.3 \\
24~$\mu$m & 0.5 & 1.3 & 2.0 \\
70~$\mu$m & 0.5 & 1.0 & 1.5 \\
100~$\mu$m & 0.7 & 1.0 & 1.7 \\
160~$\mu$m & 0.7 & 1.3 & 2.0 \\
350~$\mu$m & 1.0 & 2.0 & 3.0 \\
850~$\mu$m & 2.0 & 3.0 & 4.0 \\
1.4~mm & 2.5 & 3.5 & 4.5 \\
2.1~mm & 2.0 & 3.5 & 5.0 \\
\enddata
\end{deluxetable}

\begin{figure*}[!ht]
\begin{center}
\includegraphics[width=1.0\linewidth]{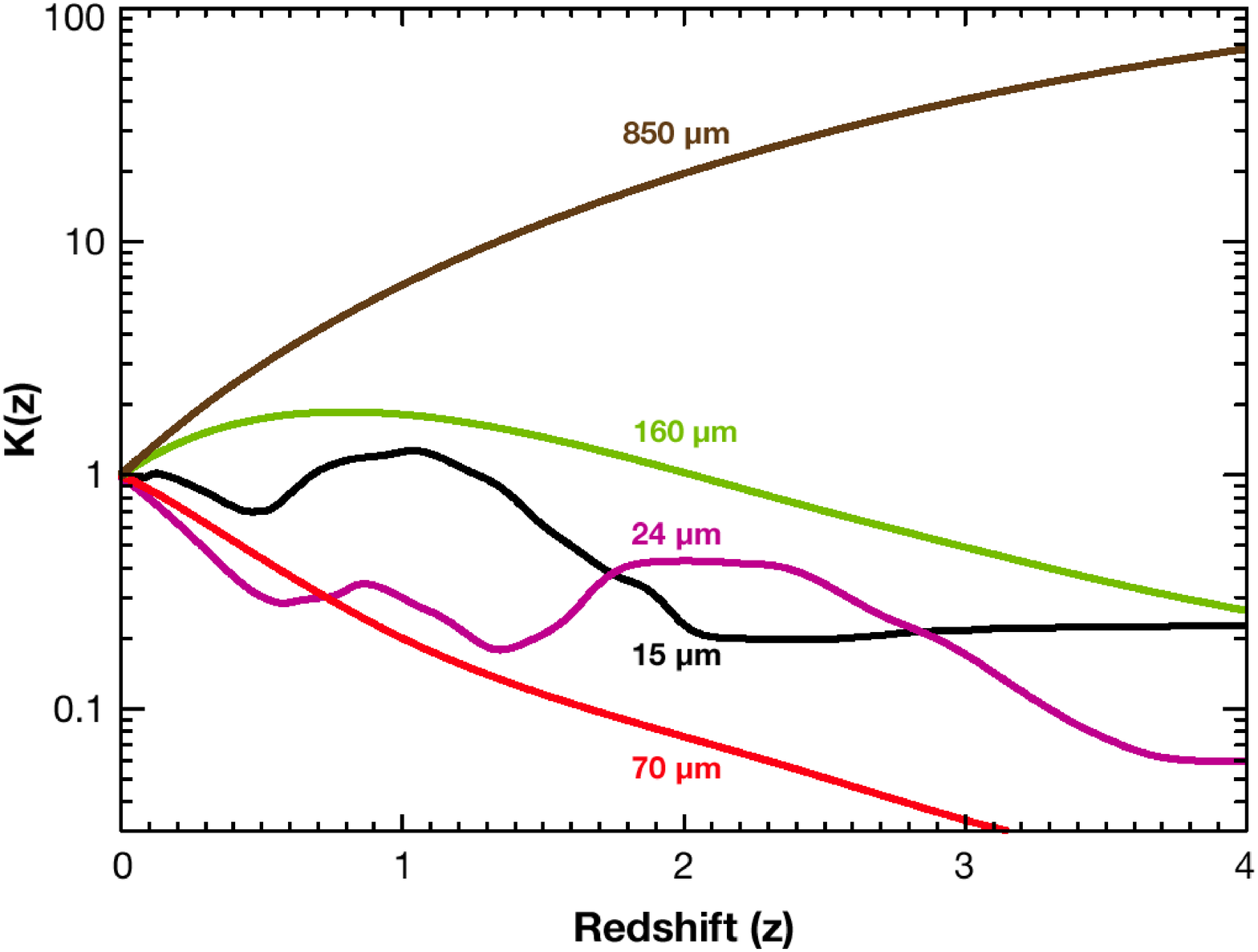}
\end{center}
\caption{K-correction at 15, 24, 70 and 160 and 850 $\rm \mu$m for a typical
LIRG with L=2$\times$10$^{11}$ L$_{\odot}$ (from the model of Lagache
et al. 2004).}
\label{fig:K_correction}
\end{figure*}

Figure \ref{fig:cib_slice_z_cumul.color.ps} shows the fraction of
resolved CIB as a function of redshift for selected wavelengths. Fifty
percents of the CIB is due to galaxies at redshift below 1 at 15 and
70~$\rm \mu$m, 1.3 at 24 and 160~$\rm \mu$m, 2 at 350~$\rm \mu$m, 3 at
850~$\rm \mu$m and 3.5 at 2000~$\rm \mu$m (see also Table
\ref{tab:cib_resolution}).  It is clear that from the far-infrared to
the millimeter, the CIB at longer wavelengths probes sources at higher
redshifts.\\

From Section \ref{surveys}, we see that the most constraining surveys
in term of resolving the CIB are those at 15, 24 and 850 $\rm \mu$m.
Moreover, the capabilities of these surveys to find high-$z$ objects
are the best among all other existing surveys (see Figure
\ref{fig:plot_lbol_vs_z4araa}).  These surveys probe the CIB in
well-defined and distinct redshift ranges, with median redshifts of
0.7 (Liang et al. 2004), $\sim$1 (Caputi et al. 2005 and L. Yan,
private communication), and 2.2 (Chapman et al. 2005) at 15, 24 and
850~$\rm \mu$m, respectively. Such well-defined redshift ranges can be
understood by looking at the K-correction. The K-correction is defined
as:
\begin{equation}
K(L,z) = \frac{L_{\nu}(1+z)}{L_{\nu}(z=0)}
\end{equation}
where $L_{\nu}(z=0)$ is the rest-frame luminosity.  This correction is
specific to the spectrum of the population considered at a given
luminosity and redshift. Figure \ref{fig:K_correction} shows the
K-correction at 15, 24, 70, 160 and 850 $\rm \mu$m. The broad plateau
observed around $z=1$ at 15~$\rm \mu$m and around $z=2$ at 24~$\rm
\mu$m is caused by the PAHs' features. At longer wavelengths, the slow
decrease of the K-correction is caused by the shape of the starburst
spectra around the peak of their emission. At 850 $\rm \mu$m, the
monotonic rise favors the detection of high-$z$ objects.

\section{GALAXIES AT REDSHIFTS $0.5 \le z \le 1.5$}
\label{zinf1}
At the time this review was written, most of the detailed informations
on dusty galaxies in the $0.5 \le z \le 1.5$ redshift range comes from
galaxies selected with the ISOCAM cosmological surveys and the
multi-wavelength analysis of detected sources. The ISOCAM
extragalactic surveys were performed with two filters, LW2 (5-8.5 $\rm
\mu$m) and LW3 (12-18 $\rm \mu$m) centered at 6.75 and 15 $\rm \mu$m,
respectively. However, because of star contamination and because the
stellar light dominates the flux from galaxies in the 6.75 $\rm \mu$m
band above redshift 0.4, only the 15 $\rm \mu$m surveys are relevant
here. Shallow, deep and ultra-deep surveys were performed in various
fields including the Lockman hole, Marano, northern and southern
Hubble Deep Field (HDF), and Canada-France Redshift Survey (CFRS)
(e.g., Aussel et al. 1999; Flores et al. 1999; Lari et al. 2001;
Gruppioni et al. 2002; Mann et al. 2002; Elbaz \& Cesarsky 2003; Sato
et al. 2003). Deeper images have been made in the direction of distant
clusters (e.g., Metcalfe et al. 2003).  Finally, the bright end of the
luminosity function was explored by the ELAIS survey (e.g., Oliver et
al. 2000).  The deepest surveys reach a completeness limit of about
100 $\rm \mu$Jy at 15 $\rm \mu$m (without lensing). The most relevant
data to this section are the deep and ultra-deep surveys.

\subsection{Detailed Properties}

To find out the nature and redshift distribution of the 15 $\rm \mu$m
deep survey sources, many followup observations have been conducted
including HST imaging and VLT spectroscopy.  With a point-spread
function full width at half of maximum of 4.6 arcsec at 15~$\rm \mu$m,
optical counterparts are easily identified. Redshifts are found using
emission and/or absorption lines. From field to field, the median
redshift varies from 0.52 to 0.8, a quite large variation due to
sample variance. Each field clearly exhibits one or two redshift
peaks, with velocity dispersion characteristic of clusters or galaxy
groups. Most of ISOCAM galaxies have redshifts between $\sim$ 0.3 and
1.2, consistent with Figure \ref{fig:plot_lbol_vs_z4araa}. About
85$\%$ of the ISO galaxies show obvious strong emission lines (e.g.,
[OII] 3723, H$_{\gamma}$, H$_{\beta}$, [OIII] 4959, 5007). These lines
can be used as a diagnostic of the source of ionization and to
distinguish the HII-region like objects from the Seyferts and LINERs.
Most of the objects are found to be consistent with HII regions, e.g.,
from Liang et al. (2004) and exhibit low ionization level
([OIII]/H$_{\beta} <$3). From emission lines studies, the AGN fraction
is quite low, $\sim$20 $\%$. This is consistent with X-ray
observations of ISOCAM sources showing that AGNs contribute at most
17$\pm$6$\%$ of the total mid-infrared flux (Fadda et
al. 2002). Assuming template SEDs typical of star-forming and
starburst galaxies, 15 $\rm \mu$m fluxes can be converted into total
infrared luminosities, L$_{\rm IR}$ (between 8 and 1000 $\rm
\mu$m). About 75$\%$ of the galaxies dominated by the star formation
are either LIRGs or ULIRGs. The remaining 25\% are nearly equally
distributed among either "starbursts" (10$^{10}<$ L$_{\rm IR}<$
10$^{11}$ L$_{\odot}$) or "normal" (L$_{\rm IR}<$ 10$^{10}$
L$_{\odot}$) galaxies. The median luminosity is about
3$\times$10$^{11}$ L$_{\odot}$. ULIRGs and LIRGs contribute to about
17$\%$ and 44$\%$ to the CIB at 15 $\rm \mu$m, respectively (Elbaz et
al. 2002). This suggests that the star formation density at $z<$1 is
dominated by the abundant population of LIRGs.  As will be shown
later, this has important consequences for the evolution of galaxies.
Because of large extinction in LIRGs and ULIRGs, the infrared data
provide more robust SFR estimate than UV tracers. The extinction
factor in LIRGs averages to A$_V$ $\sim$2.8 at $z\sim$0.7 (Flores et
al. 2004). It is much higher than that of the local star-forming
galaxies for which the median is 0.86 (Kennicutt 1992). Assuming
continuous burst of age 10-100 Myr, solar abundance, and a Salpeter
initial mass function, the SFR can be derived from the infrared
luminosities (Kennicutt 1998):
\begin{equation}
SFR (M_{\odot}/yr)=1.71 \times 10^{-10} L_{IR}[8-1000 \rm \mu m](L_{\odot})
\end{equation}
Thus typical LIRGs form stars at $\ge$20 M$_{\odot}$ year$^{-1}$.  The
median SFR for the 15 $\rm \mu$m galaxies is about 50 M$_{\odot}$
year$^{-1}$, a substantial factor larger than that found for
faint-optically selected galaxies in the same redshift range. \\

The other fundamental parameter characterizing the sources of the peak
of the infrared background is their stellar mass content that traces
the integral of the past star formation activity in the galaxies and
is a natural complement to the instantaneous rate of star
formation. The stellar masses can be obtained using spectral synthesis
code modeling of the UV-optical-near infrared data or, more simply
using the mass-to-luminosity ratio in the K-band. The derived stellar
masses for the bulk of ISOCAM galaxies range from about 10$^{10}$ to
3$\times$10$^{11} $M$_{\odot}$, compared to 1.8$\times$10$^{11}$
M$_{\odot}$ for the Milky Way. As expected from the selection based on
the LW3 flux limit -- and thus on the SFR -- masses do not show
significant correlation with redshift (Franceschini et al. 2003).  An
estimate of the time spent in the starburst state can be obtained by
comparing the rate of ongoing star formation (SFR) with the total mass
of already formed stars: $t[years]=M/$SFR.  Assuming a constant SFR,
$t$ is the timescale to double the stellar mass.  For LIRGs at
$z>$0.4, $t$ ranges from 0.1 to 1.1 Gy with a median of about 0.8 Gyr
(Franceschini et al. 2003; Hammer et al. 2005). From $z=1$ to $z=0.4$
(i.e., 3.3 Gyr), this newly formed stellar mass in LIRGs corresponds
to about 60$\%$ of the $z=1$ total mass of intermediate mass
galaxies. The LIRGs are shown to actively build up their metal
content. In a detailed study, Liang et al. (2004) show that, on
average, the metal abundance of LIRGs is less than half of the $z \sim
0$ disks with comparable brightness. Expressed differently, at a given
metal abundance, all distant LIRGs show much larger B luminosities
than local disks. Assuming that LIRGs eventually evolve into the local
massive disk galaxies, Liang et al. (2004) suggest that LIRGs form
nearly half of their metals and stars since $z\sim$1.\\

Finally, morphological classification of distant LIRGs is essential to
understand their formation and evolution. Zheng et al. (2004)
performed a detailed analysis of morphology, photometry, and color
distribution of 36 (0.4$<z<$1.2) ISOCAM galaxies using HST
images. Thirty-six percents of LIRGs are classified as disk galaxies
with Hubble type from Sab to Sd; 25\% show concentrated light
distributions and are classified as Luminous Compact Galaxies (LCGs);
22\% display complex morphology and clumpy light distributions and are
classified as irregular galaxies; only 17\% are major ongoing mergers
showing multiple components and apparent tidal tails. This is clearly
different from the local optical sample of Nakamura et al. (2004) in
the same mass range in which 27\%, 70\%, $<$2\%, 3\% and $<$2\% are
E/S0, spirals, LCGs, irregulars and major mergers
respectively. Consequences for galaxy evolution will be given in
Section \ref{scenario}. For most compact LIRGs, the color maps reveal
a central region strikingly bluer than the outer regions. These blue
central regions have a similar size to that of bulges and a color
comparable to that of star-forming regions. Because the bulge/central
region in local spiral is relatively red, such a blue core structure
could imply that the galaxy was forming the bulge (consistent with
Hammer et al. 2001). It should be noticed that they find all LIRGs
distributed along a sequence that relates their central color to their
compactness.  This is expected if star formation occurs first in the
center (bulge) and gradually migrate to the outskirts (disk), leading
to redder colors of the central regions as the disk stars were
forming.

\subsection{Cosmic Evolution}
Another remarkable property of the 15 $\rm \mu$m sources is their
extremely high rates of evolution with redshift exceeding those
measured for galaxies at other wavelengths and comparable to or larger
than the evolution rates observed for quasars. Number counts at 15
$\rm \mu$m show a prominent bump peaking at about 0.4~mJy.  At the
peak of the bump, the counts are one order of magnitude above the
non-evolution models. In fact, data require a combination of a
(1+z)$^3$ luminosity evolution and (1+z)$^3$ density evolution for the
starburst component at redshift lower than 0.9 to fit the strong
evolution. Although it has not been possible with ISOCAM to probe in
detail the evolution of the infrared luminosity function, {\it
Spitzer} data at 24~$\rm \mu$m give for the first time tight
constraints up to redshift 1.2 (Le Floc'h et al. 2005;
P\'erez-Gonz\'alez et al. 2005). A strong evolution is noticeable and
requires a shift of the characteristic luminosity L$^{\star}$ by a
factor (1+z)$^{4.0\pm0.5}$. Le Floc'h et al. (2005) find that the
LIRGs and ULIRGs become the dominant population contributing to the
comoving infrared energy density beyond $z\sim$0.5-0.6 and represent
70\% of the star-forming activity at $z\sim$1. The comoving luminosity
density produced by luminous infrared galaxies was more than 10 times
larger at $z\sim$1 than in the local Universe. For comparison, the
B-band luminosity density was only three times larger at $z=1$ than
today. Such a large number density of LIRGs in the distant Universe
could be caused by episodic and violent star-formation events,
superimposed on relatively small levels of star formation activity.
This idea has emerged in 1977 (Toomre 1977) and is fully developed in
Hammer et al. (2005).  These events can be associated to major changes
in the galaxy morphologies. The rapid decline of the luminosity
density from $z=1$ is only partially due to the decreasing frequency
of major merger events. Bell et al. (2005) showed that the SFR density
at $z\sim$0.7 is dominated by morphologically normal spiral, E/S0 and
irregular galaxies ($\ge$70\%), while clearly interacting galaxies
account for $<$30\%. The dominent driver of the decline is a strong
decrease in SFR in morphologically undisturbed galaxies.  This could
be due, for example, to gas consumption or to the decrease of weak
interactions with small satellites that could trigger the star
formation through bars and spiral arms.\\
 
Locally 0.5\% of galaxies with L$_{\rm V}>$10$^{10}$ L$_{\odot}$ have
SEDs typical of LIRGs. This changes dramatically at higher redshift:
in deep surveys, ISO detect about 15\% of the M$_B \le$-20 galaxies
(LIRGs, Hammer et al. 2005) and {\it Spitzer} detect about 30\% of
field galaxies (Starbursts and LIRGs, Bell et al. 2005).  Thus the two
populations (optical and LIR galaxies) overlap more at high $z$.

\subsection{\label{scenario} Towards a Scenario of Recent Bulge and Disk Formation in 
Intermediate-Mass Spirals}

Because a significant part of recent star formation takes place in
LIRGs, any overall picture of galaxy evolution requires a detailed
panchromatic study.  Optical/spectral properties of LIRGs are similar
to those of other galaxies and only infrared measurements are able to
describe how the star formation is distributed between the different
galaxy types. Thus a complete study has to link the star formation
revealed in the infrared to the morphological changes seen in the
optical.  This has been done by Hammer et al. (2005) using HST, ISO,
VLA and VLT observations of the CFRS. A detailed comparison of the
morphologies of distant (0.4$<$z$<$1.2) galaxies with the local
galaxies shows the complete vanishing of the LCGs in the local
Universe (by a factor $\sim$10) and the decrease of mergers and
irregulars (by a factor $\sim$4). Almost all distant galaxies have
much bluer central colors than local bulges, probably as a result of
active star formation in the 1kpc central region of most distant
spirals. This supports a relatively recent formation of bulges in many
present-day spirals.  This simultaneous changes in galaxy morphologies
and central colors of distant galaxies together with the observed
lower metallicities (Liang et al. 2004) and overall higher
star-formation rates at high $z$ are the ingredients for an updated
scenario of bulge and disk formation in spirals. Hammer et al. (2005)
propose three different phases of galaxy evolution: the
mergers/interacting, the compact galaxy and finally the growth of disk
phase. During the last 8 Gyrs, most luminous galaxies are expected to
experience a major merger that suppresses the disk as matter is
falling to the mass barycenter.  This phase is associated with short
(1 Gyr) and strong peaks of star formations.  Most of galaxies in this
phase are LIRGs. Then, the compact phase corresponds to a decrease
over 0.6-2 Gyr of the enhanced star formation due to merging.  A
significant fraction of stars form in bulges and additional occurrence
of gas infall may subsequently wrap around the bulge to form a new
disk-like component.  Finally, the star formation spreads over all the
forming disk as fed by large amounts of gas infall. In this scenario,
about half of the bulge stellar content was made earlier in their
progenitors, before the last major phase of accretion. More than a
third of the present-day stellar mass is formed at $z<$1. This
scenario is in very good agreement with the hydrodynamical numerical
simulations of Scannapieco \& Tissera (2003) in which mergers, through
secular evolution and fusions, transform galaxies along the Hubble
sequence by driving a morphological loop that might also depend on the
properties of the central potential wells, which are also affected by
mergers.  This very attractive scenario links in a simple way the
distant and local galaxies; it will be confronted to the new
panchromatic studies of {\it Spitzer} galaxies. Note that another
possibility of buildup of dense central component in disk galaxies is
internal secular evolution, as reviewed by Kormendy \& Kennicutt
(2004).

\section{GALAXIES AT REDSHIFTS z $\ge 1.5$}

Analysis of the CIB in the light of the ISO observations shows that, as we go to wavelengths 
much longer than the emission peak,
the CIB should be dominated by galaxies at higher redshifts as illustrated in Figure 
\ref{fig:CIB_fraction_spec}. 
The comoving infrared production rate needed to fill the CIB around
1mm at a redshift centered around 2.5 to 3 remains comparable to the
one from galaxies detected in the ISOCAM surveys and filling 65\% of
the peak of the CIB.  In this section we discuss the rapidly growing
observational evidence that this picture is basically correct.  The
main source of these observations has been the SCUBA submillimeter
observations at 850~$\mu$m and 450~$\mu$m (see Blain et al 2002 for a
review) and observations from the MAMBO instrument on the IRAM 30-m
telescope at 1.2 mm (Greve et al. 2004).  The negative K-correction
becomes very effective at these wavelengths, leading to an almost
constant observed flux for galaxies of the same total infrared
luminosity between redshifts 1 and 5.  More recently, the {\it
Spitzer} observatory has produced a wealth of early observations
showing that this observatory will contribute much to our
understanding of infrared galaxies at z$\ge$1.5.

\begin{figure*}[!ht]
\begin{center}
\includegraphics[width=0.85\linewidth]{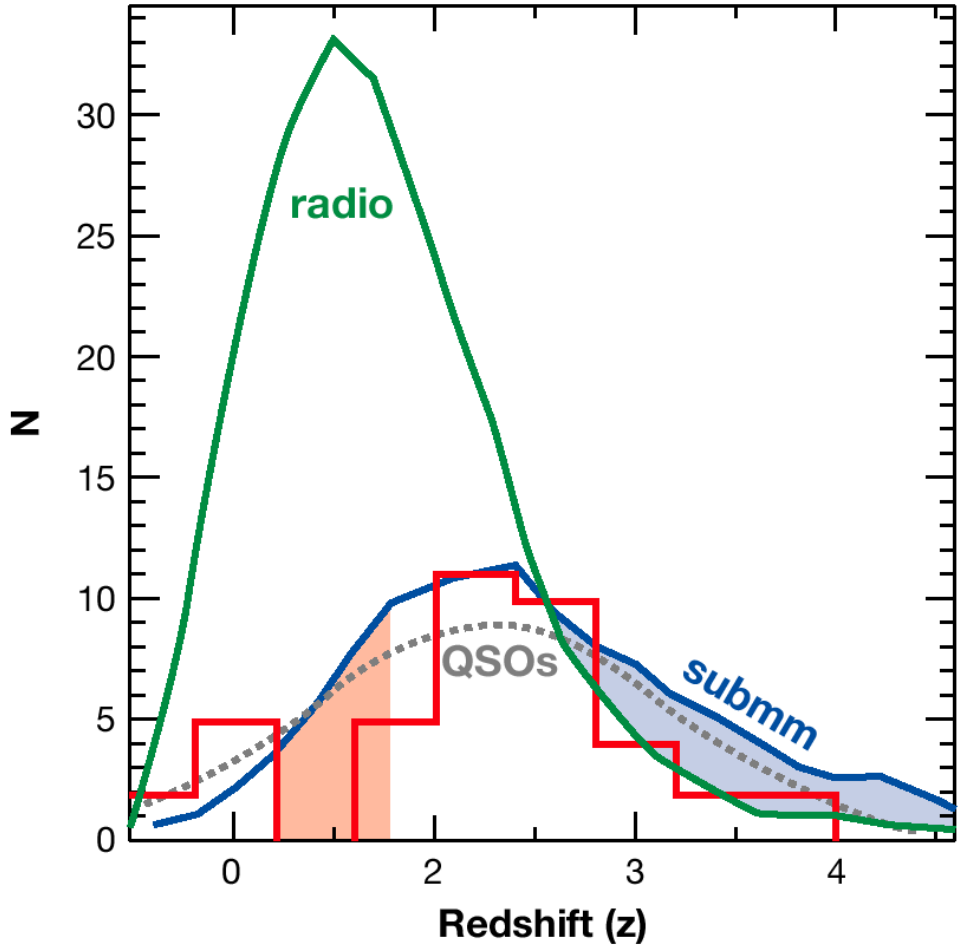}
\end{center}
\caption{The observed histogram of the redshift distribution for the 55
radio-identified SCUBA galaxies ({\it red histogram}). Curves derived for
a model of the radio/submillimeter galaxy populations (Chapman et al. 2003b; Lewis et al.
2005) are overplotted
suggesting that the redshifts
of the sources missed in the radio identification process lie mostly 
at redshifts $\sim$3-5
between the radio and submillimeter model tracks.
A sample of radio-selected QSOs is also overplotted ({\it gray dotted line}), 
revealing a remarkable
similarity with the observed distribution for submillimeter galaxies. From Chapman et al. 2003a.}
\label{fig:chapman_z}
\end{figure*}

\subsection{\label{counts} Number Counts, Contribution to the CIB}
Blank-field deep surveys combined with mapping of areas lensed by clusters lead to 
number counts at 850 $\mu$m down to 0.5 mJy (e.g., Smail et al. 2002; Wang et
al. 2004). 
At 1.2~mm counts have been obtained down to 2.5 mJy (e.g., Greve et al. 2004). The number counts shapes
at 850~$\mu$m and 1.2~mm are compatible with the assumption that they are made of 
the same population 
with a flux  ratio $F_{850} / F_{1200} = 2.5$. 
For a typical ULIRG SED, a 5mJy source at 850~$\mu$m has a luminosity 
of $10^{12} L{\odot}$ at a redshift of about 2.5.
The large fraction of the background resolved at 850~$\mu$m (see
Section \ref{surveys}) has interesting consequences. It shows very
directly that if the sources are at redshift larger than 1 (as
confirmed by the redshift surveys discussed below), the infrared
luminosity of the sources that dominate the background is larger than
$10^{12} L_{\odot}$. This is a population with a very different
infrared luminosity function than the local or even the $z=1$
luminosity function. The link between this population at high $z$, and
what has been seen at $z \sim 1$ (as discussed in Section \ref{zinf1})
will be done by {\it Spitzer}/MIPS observations at 24~$\mu$m.  Figure
\ref{fig:CIB_fraction_spec} shows that the building on the bulk of the
CIB near its peak (at 150~$\mu$m) with redshift is expected to be
similar to the building of the 24~$\mu$m background when the history
of the 15 and 70~$\mu$m CIB have larger contributions from redshift-1
sources.  The K-correction plots (Figure \ref{fig:K_correction}) show
for 15 $\mu$m a hump at $z=1$ associated with the coincidence of the
6-9 $\mu m$ aromatic features in the ISOCAM filter and a hump at the
same redshift for the 24~$\mu$m MIPS filter associated with the
11-14~$\mu$m set of aromatic features in the MIPS filter.  For the
MIPS filter a second hump is visible at $z \sim$2 that corresponds to
6-9~$\mu$m features centered on the 24~$\mu$m MIPS filter. ISOCAM
galaxies contribute to about 2/3 of the energy peak of the CIB.
Following the previous considerations, it is easy to understand why
the remaining fraction is likely to be made of sources in the redshift
range 1.5-2.5. The presently detected SMGs with luminosity $10^{12}
L_{\odot}$ have an almost constant flux between redshift 1.7 and
redshift 2.5 at 24~$\mu$m (similar to the constant flux at 850~$\mu$m
between redshift 1 and 5). The MIPS 24~$\mu$m deep surveys (e.g.,
Papovich et al. 2004) reach a sensitivity of 50 $\mu$Jy and thus can
detect all these galaxies when they are starburst-dominated.
Considering the speed of the MIPS it is likely that 24~$\mu$m surveys
will become the most efficient way to search for luminous starburst
galaxies up to $z= 2.5$ and up to 3 for the most luminous ones.

\subsection{\label{redshifts-dist} Redshift Distribution and SEDs of the SMGs}
The first obvious question when investigating the nature of the
submillimeter galaxies (SMGs) is their redshift distribution. The
rather low angular resolution of the submillimeter and millimeter
observations made identifications with distant optical galaxies an
almost impossible task without an intermediate identification. This is
provided by radio sources observed with the VLA with 10 times better
angular resolution. The tight correlation between far-infrared
luminosity and radio flux (Helou et al. 1985; Condon 1992) provides
the needed link.  This then allows us to get optical identifications
and redshift measurements using 10-m class telescopes.  Confirmation
of these identifications can then be obtained through CO line
observations with the millimeter interferometers such as the Plateau
de Bure interferometer.  The redshift deduced from the optical lines
is confirmed by the CO observations.  So far, only a handful of cases
have gone through this whole chain of observations (e.g., Genzel et
al. 2003; Greve et al. 2005; Neri et al. 2003), but a high success
rate gives confidence in the first step of the identification
process. The chain will also have to be applied to the tentative
counterparts of radio-undetected SMGs that have been found using a
certain combination of optical properties (Pope et al. 2005).

The difficulty of making large, blind surveys at 850~$\mu$m at the
required sensitivity has lead to an attempt to find distant SMGs
through blind surveys at different wavelengths. Barger et al. (2000)
have observed optically faint radio sources at submillimeter
wavelength and demonstrated them to be, so far, the most efficient way
to preselect targets for submillimeter observations and to get larger
samples of potentially high redshift SMGs.  As an example, Chapman et
al. (2002) recovered at 850~$\mu$m 70\% of the blind submillimeter
survey sources. This contrasts the recovery rate of MAMBO sources,
which is relatively low, $\sim$25\% (Dannerbauer et al. 2004).  It
should be noticed that the radio preselection biases the sample
against very high redshifts ($z<$3) because the radio flux at 1.4 GHz
is below the detection limit of the VLA surveys used for this
preselection.  A model by Chapman et al. (2003b) and by Lewis et
al. (2005) illustrate this effect very well (Figure
\ref{fig:chapman_z}). A fraction of the submillimeter-selected sources
are missed in such a process at $z>$ 3 (detectability in radio) and
around $z\sim$1.5 (optical redshifts desert). The number of non
identified submillimeter sources (around 30$\%$ for $S_{850}>3$ mJy)
is consistent with this model. Nevertheless the submillimeter-selected
sources do not appear qualitatively different from the
optically-faint-radio selected ones.  Another bias is the effect of
the dust effective temperature of the SMGs (Lewis et al. 2005). At a
given total far-infrared luminosity, hotter sources have lower
submillimeter fluxes if the radio/far-infrared correlation continues
to hold. They could be missed in the submillimeter surveys (see the
discussion in Chapman et al. 2005).
\\

\begin{figure*}[!ht]
\begin{center}
\includegraphics[width=1.0\linewidth]{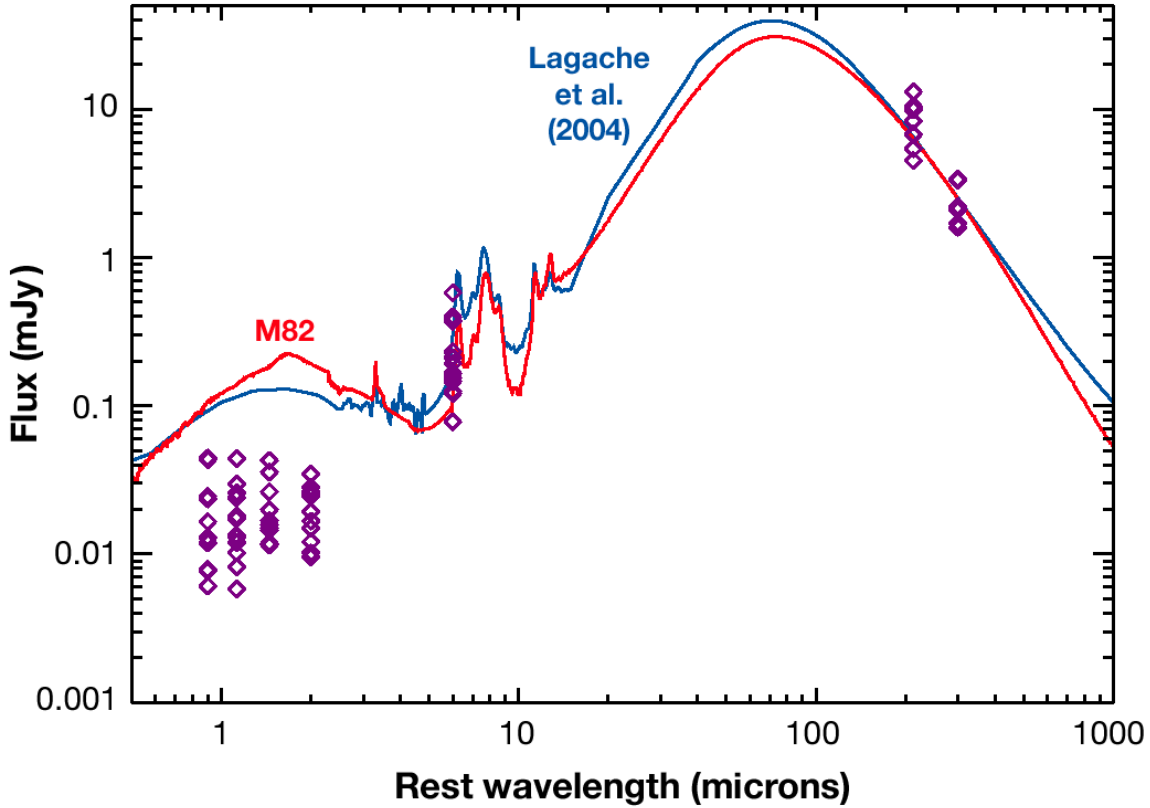}
\end{center}
\caption{Rest-frame SED of 15 SMGs (assuming a redshift of 3)
with MAMBO and/or SCUBA, {\it Spitzer}/IRAC
and {\it Spitzer}/MIPS 24 $\mu$m measurements. {\it Purple diamonds} are the galaxies 
208, 119, 115, 48, 44 (Frayer et al.
2004), LE850$\_$4, LE850$\_$35 (Egami et al. 2004), and
MMJ105201, MMJ105155, MMJ105203, MMJ105216, MMJ105148, MMJ105157, MMJ105207
MMJ105203 (Ivison et al. 2004). Overplotted are the SEDs of M82, 
normalized at 850~$\mu$m (from Chanial 2003), and the SED template of 
the Lagache et al. (2004) model, for L=10$^{13}$ L$_{\odot}$ and a 
redshift of 3 (no normalization has been applied).
Note that this sample of SMGs has a ratio dust/stellar component higher
than the template or M82.
}
\label{fig:SED}
\end{figure*}

Chapman et al. (2003a) got spectroscopic redshifts of 55 sources
obtained in this way.  The redshift distribution for these sources is
shown in Figure \ref{fig:chapman_z} (note that when this review was
being edited, Chapman et al. (2005) publish spectroscopic redshifts
for 73 SMGs). This distribution peaks at z= 2.4 with a substantial
tail up to z=4. In fact the redshift distribution can be represented
by a Gaussian distribution centered on 2.4 and with a sigma of
0.65. Almost all SMGs are found in the redshift range $1.5<z<3$. This
redshift distribution is compared with that of the redshift
distribution for a pure radio sample in Figure \ref{fig:chapman_z}.
The SMGs selected in the way described above is also shown to be very
similar to the redshift distribution of the radio-selected QSOs. This
observation is interesting in the context of high rate of AGN activity
detected in SMGs.
\\
The determination of the SED of millimeter/submillimeter galaxies
remains an open question despite a lot of work in the last few
years. The SCUBA and MAMBO data provide constraints on the flux and
spectrum at long wavelengths; {\it Spitzer} observations constrain the
near and mid-infrared. The far-infrared part of the SED remains the
least precisely known. Low angular resolution makes 70 and 160~$\mu$m
deep surveys confusion-limited at 3 and 40 mJy (Dole et
al. 2004b). These limits are too high to complete the SED of the SMGs
(see Figure \ref{fig:SED}).  Stacking sources will help to go deeper
than the confusion limit when large samples of SMGs are available in
MIPS cosmological surveys. A first attempt on a radio-selected sample
lowered the limit down to 1.2~mJy at 70~$\mu$m (Frayer et al. 2004).
They find a typical flux ratio I(70)/I(24)$<$7 that they interpret to
be low when compared with low-redshift starburst. However, such low
ratios are typical of dusty starbursts placed at redshift greater than
1.5. It is thus likely that the lower colors are due to a redshift
effect.  Appleton et al. (2004) looked at the mid- and far-infrared
fluxes from a purely radio-selected 1.4 GHz $\mu$Jy sample of about
500 and 230 sources at 24 and 70~$\mu$m, respectively.  They show that
the far-infrared to radio correlation that is constant out to $z=1$
seems to be constant using 24~$\mu$m out to $z=2$ but with a larger
dispersion due systematic variations in SED shape throughout the
population. This provides positive evidence of the universality of the
infrared/radio correlations out to redshifts of about 2.
\\

\begin{figure*}[!ht]
\begin{center}
\includegraphics[width=0.85\linewidth]{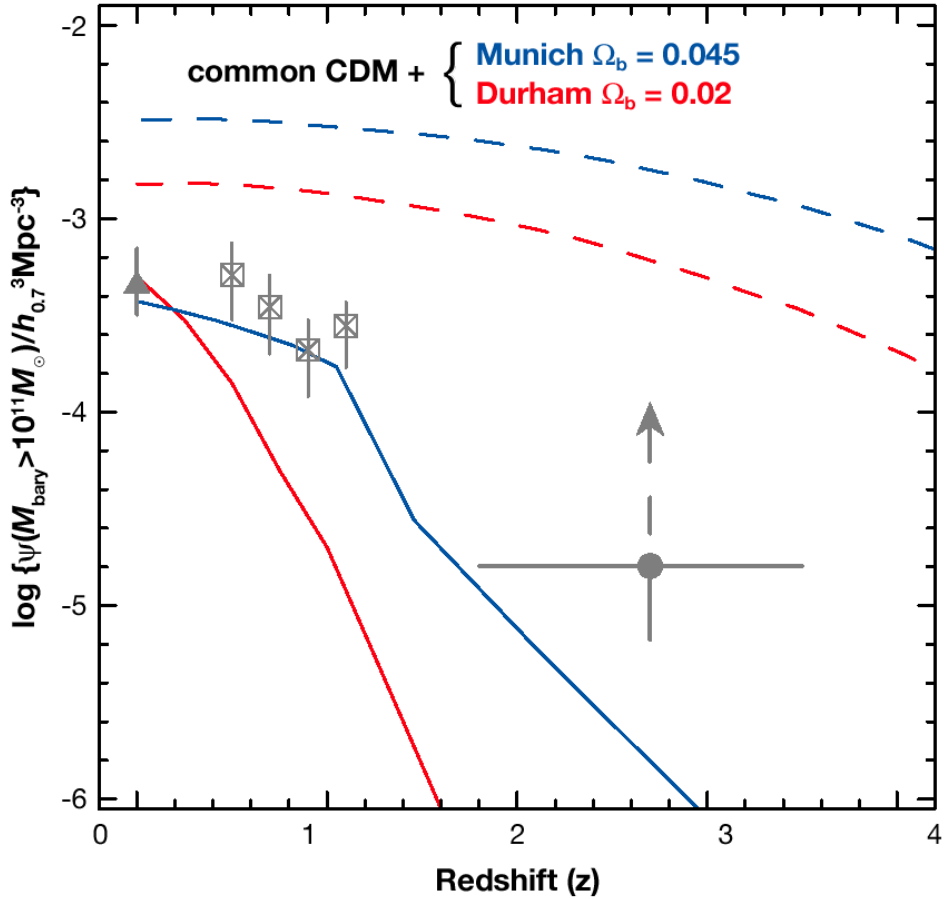}
\end{center}
\caption{Comoving number densities of galaxies with baryonic masses $\geq
10^{11}\,M_\odot$ as a function of redshift.  The {\it triangle} and {\it open 
squares} show densities of massive stellar systems at $z = 0$ 
and $z \sim 1$; The {\it circle} shows the density for massive SMGs at 
$z \sim 2.7$, with a factor of 7 correction for burst lifetime. {\it Blue}
and {\it red curves}
show the predictions of semianalytic models by the ``Munich'' and 
``Durham'' groups, respectively.  {\it Dashed curves} show the corresponding 
number densities of halos with available baryonic masses 
$\geq 10^{11}\,M_\odot$.  The two models use the same halo simulations but 
assume different $\Omega_{\rm b}$. From Genzel et al. (2004).}
\label{fig:genzel}
\end{figure*}

Blain et al (2004a) have analyzed SEDs of infrared galaxies assuming
that the low-redshift radio/far-infrared correlation applies to SMGs.
Under this reasonable assumption and using a model of long-wavelengths
SEDs based on a single modified black body, they can choose a single
parameter to built an SED that fits the long-wavelengths data and the
radio/infrared luminosity ratio. In their paper, this single parameter
is the temperature, but it could equally well be the long-wavelength
emissivity, because they showed that this is degenerate with
temperature.  A split between two redshift populations appear in their
analysis. The high-$z$ galaxies selected by the submillimeter
observations are significantly colder that the low-$z$ galaxies (Dune
\& Eales, 2001; Stanford et al. 2000), IRAS or IRAS-radio
selected. The discrepancy in part probably reflects selection effects
in the way these samples were obtained and may reflect the fact that
SMGs and local infrared galaxies are distinct populations.  It remains
an open question what effect this has on the SED model. The main worry
is that a single modified black body often does not fit ULIRGs SED
when they are known at many frequencies.  The SED is broader; the
unavoidable temperature distribution of dust in infrared galaxies
would affect such an analysis. In fact, the Stanford et al. (2000)
sample does not agree well with the single-temperature SED, and this
led Lagache et al. (2004) to take broader SEDs for their starburst
galaxy templates.

\subsection{\label{nature of SMG} Nature of the SMGs}

Many LIRGs and ULIRGs at low redshifts have been identified with
interacting or galaxy mergers. A substantial fraction show signs of
AGN activity but it has been shown for the low-redshift LIRGs and
ULIRGs that the starburst component dominates the energy output
(Genzel et al. 1998; Lutz et al. 1998). The sources used for the
redshift distribution by Chapman et al. (2003a) have been imaged with
the HST.  Most of them are multi-component-distorted galaxy systems
(Conselice et al. 2003; Smail et al. 2004). They display irregular and
frequently highly complex morphologies compared to optically selected
galaxies at similar redshifts. They are often red galaxies with bluer
companions, as expected for interacting, star-forming galaxies.  They
have higher concentrations, and more prevalent major-merger
configurations than optically-selected galaxies at $z\sim$2-3.  Most
strikingly, most of the SMGs are extraordinarily large and elongated
relative to the field population regardless of optical magnitude
(Chapman et al. 2003c). SMGs have large bolometric luminosities, $\sim
10^{12}-10^{13}$ L$_{\odot}$, characteristic of ULIRGs. If the
far-infrared emission arises from the star formation, the large
luminosities translate to very high SFR $\ge$1000 M$_{\odot}$
year$^{-1}$. Such high rates are sufficient to form the stellar
population of a massive elliptical galaxy in only a few dynamical
times, given a sufficient gas reservoir.  SMGs are very massive
systems with typical mass of 1-2$\times$10$^{11}$L$_{\odot}$ (Swinbank
et al. 2004), comparable to the dynamical mass estimates from CO
observations.  Genzel et al. (2004; and more recently Greve et
al. 2005) have undertaken an ambitious program to study the nature of
the SMGs in more details. They got CO spectra with the Plateau de Bure
interferometer for 7 sources out of their sample of 12 for the CO 3-2
and 4-3 transitions redshifted in the 3~mm atmospheric window.  They
provide optical identifications and redshifts.  The detection of these
sources at the proper redshift confirms the usefulness of
identification with the help of the radio sources.  The median
redshift of this sample is 2.4.  In addition, one source was studied
with the SPIFI instrument on the ESO/VLT.  These observations are
giving very interesting clues on the nature of the submillimeter
galaxies.  The gas masses obtained for these systems using CO
luminosity/mass of gas determined from local ULIRGs is very large with
a median of 2.2$\times$$10^{10} M_{\odot}$ (10 times larger than in
the Milky Way).  Using the velocity dispersion, they could infer that
the dynamical median mass of these systems is 13 times larger than in
Lyman-break galaxies (LBGs) at the same redshift or 5 times the mass
of optically selected galaxies at this redshift.  These SMGs with a
flux at 850~$\mu$m larger than 5~mJy are not very rare and unusual
objects, because they contribute to about $20 \%$ of the CIB at this
frequency.  Through multiwavelength observations, Genzel et al. (2004)
get the stellar component in K band, and infer the star-formation rate
and duration of the star-formation burst. They can then compare the
number density of these massive systems with semiempirical models of
galaxy formation. The very interesting result is that this number
density is significantly larger than the predicted one, although the
absolute numbers depends on a number of assumptions like the IMF. The
comparison is shown in Figure \ref{fig:genzel}.  Such massive systems
at high redshift are not easy to understand in current cold dark
matter hierarchical merger cosmogonies. However, one must keep in mind
that bright SMGs (S$_{850}>$5~mJy) that contribute $20 \%$ of the CIB
may not be representative of the whole population. Gravitational lens
magnification provides a rare opportunity to probe the nature of the
distant sub-mJy SMGs. Kneib et al. (2005) study the property of one
SMG with an 850~$\mu$m flux S$_{850}$=0.8 mJy at a redshift of
$z=2.5$. This galaxy is much less luminous and massive than other
high-$z$ SMGs. It resembles to similarly luminous dusty starbursts
resulting from lower-mass mergers in the local Universe.\\

In order to link the different population of high-redshift objects,
several LBGs at redshift between 2.5 and 4.5 have been targeted at
850~$\mu$m. The Lyman-break technique (Steidel et al. 1996) detects
the rest-frame 91.2 nm neutral hydrogen absorption break in the SED of
a galaxy as it passes through several broad-band filters. LBGs are the
largest sample of spectroscopically confirmed high-redshift
galaxies. Observing LBGs in the submillimeter is an important goal,
because it would investigate the link, if any, between the two
populations. However, the rather low success rate of submillimeter
counterpart of LBGs (e.g., Chapman et al. 2000; Webb et al. 2003)
argues against a large overlap of the two populations.

\subsection{\label{Spitzer 24} {\it Spitzer} 24~$\mu$m Sources}
A potential new way to find high-$z$ LIRGs and ULIRGs appeared
recently with the launch of the {\it Spitzer}
observatory. Particularly suited to this goal is the 24~$\mu$m channel
of the MIPS instrument.  The confusion levels in the 70 and 160~$\mu$m
prevent detection a significant number of high-redshift objects, and
the IRAC 3.6 to 8~$\mu$m at high redshift probes mostly the old
stellar component that is much weaker than the dust emission in
starburst galaxies.  At the time of writing, the observations are
under way, and only a few results are available. Le Floc'h et
al. (2004) give the first hint on the 24~$\mu$m selected
galaxies. They couple deep 24~$\mu$m observations in the Lockman hole
and extended groth strip with optical and near-infrared data to get
both identification and redshift (either spectroscopic or
photometric).  They find a clear class of galaxies with redshift 1$\le
z \le$2.5 and with luminosities greater than $\sim$5$\times$10$^{11}$
L$_{\odot}$ (see also Lonsdale et al. 2004). These galaxies are rather
red and massive with M$>$2$\times$10$^{10}$ M$_{\odot}$ (Caputi et
al. 2005). Massive star-forming galaxies revealed at $2\leq z \leq 3$
by the 24~$\mu$m deep surveys are characterized by very high star
formation rates -- SFR $\geq$500~M$_{\odot}$ year$^{-1}$.  They are
able to construct a mass of $\simeq$10$^{11}$~M$_{\odot}$ in a burst
lifetime ($\simeq$0.1 Gyr). The 24~$\mu$m galaxy population also
comprises sources with intermediate luminosities (10$^{10} \leq L_{IR}
\leq$10$^{11}$ L$_{\odot}$) and low to intermediate assembled stellar
masses (10$^{9} \leq$M$\leq$10$^{11}$ M$_{\odot}$) at $z\leq0.8$. At
low redshifts, however, massive galaxies are also present, but appear
to be building their stars quiescently in long timescales (Caputi et
al. 2005).  At these redshifts, the efficiency of the burst-like mode
is limited to low mass M$\leq$10$^{10}$ M$_{\odot}$ galaxies. These
results support a scenario where star-formation activity is
differential with assembled stellar mass and redshift, and proceed
very efficiently in massive galaxies (Caputi et al. 2005).

In the Lockman Hole, only one galaxy is associated with an X-ray
source. This suggests that these galaxies are mostly dominating by
star formation, consistent with the findings of Alonso-Herrero et
al.(2004) and Caputi et al. (2005). This is also suggested by SEDs
that are best fitted by PAH features rather than by strongly rising,
AGN-type continua (Elbaz et al. 2005).  The selected sources exhibit a
rather wide range of MIPS to IRAC flux ratio and optical/near-infrared
shapes, suggesting a possibly large diversity in the properties of
infrared galaxies at high redshift as noticed by Yan et
al. (2004b). Based on these first analyzes, together with the
interpretation of the number counts (e.g., Lagache et al. 2004), it is
clear that the 24~$\mu$m observations will provide the sample to
unambiguously characterize the infrared galaxies up to
$z\simeq2.5$. They should fill the gap between the ISO- and
SCUBA-selected galaxies.\\

Several 24~$\mu$m observations have been conducted on selected ERO and
SCUBA and MAMBO samples. To our knowledge, LBGs have not been observed
at long wavelengths. The MAMBO/SCUBA selected galaxies in the Lockman
hole with radio identification have been observed by {\it Spitzer} and
most of them detected between 3.6 and 24~$\mu$m. This allows to get an
average SED for these (Egami et al. 2004; Ivison et al. 2004; see
Figure \ref{fig:SED}) {\it Spitzer} deep surveys at 24~$\mu$m and
shallow surveys like the SWIRE legacy (Lonsdale et al. 2004) can
easily detect them and are thus a promising new way to find this class
of high-$z$ infrared galaxy. Nevertheless, the Early Release
Observations from {\it Spitzer} have been used to extract their
submillimeter flux from a stacking analysis of SCUBA observations in
the Lockman hole (Serjeant et al. 2004). In this field, seven SMGs
were already known and others were identified by further analysis. For
the bulk of the 24~$\mu$m sources a marginal detection is found with
an $S_{850}/S_{24}$ ratio (1/20) much lower than that observed for
SMGs. This clearly shows that the SMGs are only a fraction of the
24~$\mu$m sources, as expected. An interesting challenge is to find if
{\it Spitzer} color criteria can be found to extract preferentially
SMGs, i.e., the galaxies that account for most of the CIB near
1~mm. The SED in the thermal infrared appears quite variable for LIRGs
and ULIRGs making this difficult (e.g., Armus et al. 2004).\\

Extremely Red Objects (EROs) are usually selected based on their red
colors: $(R-K_s)\ge$5.3 mag or $(I-K_s)\ge$4 mag.  This color
selection should include early-type galaxies at $z\sim$1. However, the
color selections are also sensitive to dust-reddened, star-forming
systems. Up to now, it remains unclear what fraction of EROs are truly
dust-obscured galaxies. Different scenarios of galaxy formation
predict very different formation epochs for such galaxies.  It is thus
interesting to characterize these galaxies, in particular whether they
belong to the early-type or dusty star-forming class of objects.  {\it
Spitzer}/MIPS 24~$\mu$m observations offer the first opportunity to
address this issue because 24~$\mu$m observations can clearly
discriminate between the two populations. In the N1 field, Yan et
al. (2004a) suggest that about 50\% of EROs are infrared luminous,
dusty starbursts at $z\ge1$ (in a similar study, Wilson et al. (2004)
show that at least 11\% of $0.6<z<1.3$ EROs and at least 22\% of
$z>1.3$ EROs are dusty star-forming galaxies). Their mean 24~$\mu$m
flux corresponds to infrared luminosities of about 3$\times$10$^{11}$
and 10$^{12}$ L$_{\odot}$ at z$\sim$1 and $z\sim$1.5, respectively.
They are massive galaxies with lower limit M$\geq$5$\times$10$^9$ to 2
10$^{10}$ M$_{\odot}$.  The fraction of EROs likely to be AGN is
small; about 15$\%$.  The link between the two classes of EROs could
be that starburst EROs are experiencing, at $z>1$, violent
transformations to become massive early-type galaxies.

\subsection{\label{AGN} ULIRGs and Active Galactic Nuclei at High Redshifts}

SMGs are massive ULIRGs at high redshift. One of the key question
discussed above for the $z\simeq $1 galaxies is to distinguish whether
starburst or AGN activity powers the dust heating and associated
infrared emission.  The presence of an AGN in galaxies can be
investigated using optical/near-infrared, emission line diagnostics
and/or X-ray observations.  But the identification of the presence of
an AGN does not mean that it is the dominant source of the
far-infrared emission. Alexander et al. (2003; see also Almaini et
al. 2003) use Chandra observations of the CDF-N to constrain the X-ray
properties of 10 bright SMGs. Half of the sample has flat X-ray
spectral slopes and luminous X-ray emission, suggesting obscured AGN
activity. However, a comparison of the AGN-classified sources to the
well-studied, heavily obscured AGN NGC 6240 suggests that the AGN
contributes on average a negligible fraction (about 1.4$\%$) of the
submillimeter emission.  For the MAMBO sources, similar results are
found: only one out of the nine MAMBO sources studied by Ivison et
al. (2004) has an X-ray counterpart. It has, as expected from low
redshift ULIRGs observations (e.g., Rigopoulou et al. 1999), a
different mid-infrared SED than the starburst dominated sources.
About 75\% of their sample has rest-frame mid-infrared to far-infrared
SED commensurate with obscured starburst.  Swinbank et al. (2004),
using AGN indicators provided by near-infrared spectra, estimate that
AGNs are present in at least 40$\%$ of the galaxies in their sample of
30 SMGs. Emission-line diagnostics suggest that star formation is the
dominant power source. However, the composite spectrum for the
galaxies that individually show no signs of an AGN in their
near-infrared spectra appears to show an underlying broad H$_{\alpha}$
line. This suggests that even these galaxies may host a low-luminosity
AGN that is undetectable in the individual spectra.  All these studies
tend to show that starburst activity is the dominant source of power
of dust emission in the far-infrared. Still, it is rather difficult to
estimate the true ``contamination'' by the AGN.  To go deeper, Chapman
et al. (2004) tried an original approach. They observe a sample of
identified SMGs at high angular resolution in the radio and use the
radio emission as a proxy for the far-infrared emission.  This
assumption is based on the well-known very tight far-infrared/radio
correlation mentioned above. If detected, an extended radio (and thus
far-infrared) component is likely to arise from the star
formation. The detection of extended emission requires sub-arcsec
resolution to map emission on kpc-scales. These are accessible by
radio interferometry (they are well beyond far-infrared and
submillimeter facilities capabilities). They find that for 70\% of the
SMG sample, the MERLIN/VLA radio exhibits resolved radio emission
which mirrors the general form of the rest frame UV morphology seen by
HST. The galaxies are extended on scales of about 10 kpc.  They
interpret this as a strong support for the hypothesis that radio
emission traces spatially extended massive star formation within these
galaxies.  This is clearly different from what is seen in local ULIRGs
where the far-infrared/radio emission is concentrated in the compact
nuclear region with an extend less than 1 kpc.  In the remaining 30\%
of the SMG sample, the radio emission is more compact (essentially
unresolved). This is a signature of either a compact nuclear starburst
and/or an AGN.
\\
\begin{figure}[!ht]
\begin{center}
\includegraphics[width=1.0\linewidth]{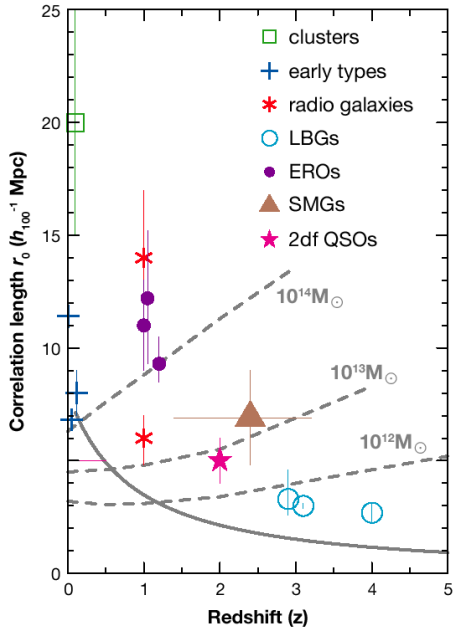}
\end{center}
\caption{Comoving correlation length of the SMGs ({\it triangle}) in 
contrast to other populations of low- and high-redshift galaxies (see
the summary in Overzier et al. 2003). The horizontal error bar on the 
SMG point spans the 
range of redshifts over which SMG associations are found. 
The {\it solid line} shows 
a representative model for 
the evolution of a certain overdensity. The {\it dashed 
lines} show the expected correlation length of dark 
matter halos as a function of mass and redshift. From Blain et al. (2004a).}
\label{fig:blain}
\end{figure}

In conclusion, the exact fraction of distant submillimeter and
millimeter galaxies containing an energetically dominant AGN is
difficult to extract from observations.  However, even in the systems
containing an unambiguously powerfully AGN, the far-infrared emission
seems to be powered by the star formation. Surprisingly, this seems to
be also the case in distant QSOs. Recently, Beelen (2004) has shown
that the far-infrared and blue luminosities from the host galaxies of
distant radio-quiet QSOs, are slightly correlated. The far-infrared
and radio emission of these quasars follow the radio-infrared
correlation observed in local ULIRGs (Yun et al. 2001), providing a
first indication that the dust is predominantly heated by the
star-formation activity rather than by the AGN. Moreover, the
non-linearity between the far-infrared and blue luminosities is also
an indication that the heating mechanism of the dust is not directly
linked to the AGN. However, the presence of this correlation could
suggest a causal connection between the formation of stars in the host
galaxy and the activity of the central super massive black hole. This
connection has been successfully modeled by Granato et al. (2004).
\\
Finally Houck et al. (2005) and Yan et al. (2005) demonstrate the
potential of using mid-infrared spectroscopy, especially the aromatic
and silicate features produced by dust grains to directly probe
distant L$\sim$10$^{13}$ L$_{\odot}$ ULIRGs at $z\sim2$. {\it
Spitzer}/IRS observations provide a unique and direct access to
high-$z$ ULIRG physical properties. It will definitively open the
route toward a complete census of the distant infrared-luminous
Universe.  A first study on two distant SMGs using {\it Spitzer}/IRS
by Lutz et al. (2005) finds for one SMG an equal contribution from
star formation and AGN. The second galaxy is dominated by star
formation.

\section{CLUSTERING}
Measuring clustering gives information about the distribution of
galaxies with respect to the dark matter. The strength of clustering
is correlated with the mass of extended halos that host luminous
galaxies.  At redshift lower than 1.3, large-scale structures have
been mapped by the DEEP2, 2dF and SDSS surveys (Coil et al. 2004;
Peacock et al. 2001; Doroshkevich et al. 2004). At higher redshifts,
correlation lengths $r_0$ have been measured for different galaxy
populations such as the LBGs and EROs. Values of about $r_0$=4
h$^{-1}$ Mpc (Porciani \& Giavalisco 2002) and $r_0$=11 h$^{-1}$ Mpc
(Daddi et al. 2000; Firth et al. 2002; Roche et al. 2003) are reported
with mean redshift of about 3 and 1-1.5 for the LBGs and EROs,
respectively.  Getting information on the clustering of the infrared
galaxies is essential to see how they relate to the other galaxy
populations and to understand their formation process.  As an example,
one of the key questions is to see if the most massive SMGs are
associated with the most massive dark matter halo.
\\

Up to now, very little information is available on the clustering of
infrared galaxies. A clustering signal can tentatively be measured,
although the small number of objects in the deep narrow-pencil ISOCAM
and SCUBA beams prevents accurate measurements of the autocorrelation
function.
At $z\sim$1, Moy \& Elbaz (2005) study the large-scale and close
environment of ISOCAM galaxies. They find that infrared galaxies are
more strongly clustered than optical galaxies. Eighty percents of
ISOCAM galaxies are found preferentially in redshift peaks, versus
68$\%$ for the optically selected galaxies. Moy \& Elbaz (2005) find
indirect evidences that the triggering mechanism of dusty starbursts
is small-scale ($\sim$ 100 kpc) galaxy-galaxy interactions.  Such
interactions do not lead to major mergers most of the time but are
more likely simple fly-by, tidal interactions or minor mergers. At
much higher redshift, there are some indirect evidences of strong
clustering of SMGs when compared to other classes of high-redshift
galaxies.  For example, Chapman et al. (2001) identified SMGs within
the most overdense structure of LBGs at $z\sim$3.1 (Steidel et
al. 2000).  De Breuck et al. (2004) detected an overdensity of MAMBO
sources likely at $z\simeq$4.1 in a proto-cluster containing also
overdensities of Ly$\alpha$ emitters and LBGs. The first
three-dimensional quantitative measurement of the clustering strength
of SMGs has been made by Blain et al. (2004a). They find spectroscopic
evidence for clustering.  Using 73 spectroscopically identified
galaxies, they find a surprisingly large number of ``associations''
with redshifts separated by less than 1200 km s$^{-1}$. They provide
tentative evidence for strong clustering of SMGs at $z\simeq$2-3 with
a correlation length of $\sim$(6.9$\pm$2.1) h$^{-1}$ Mpc using a
simple pair-counting approach appropriate for the small and sparse SMG
sample. This correlation length appears to be somewhat larger than
that for both LBG and QSO galaxies at comparable redshifts. It is thus
unlikely that the SMGs form a simple evolutionary sequence with either
population. On the contrary, the correlation length could be
consistent with a form of evolution that subsequently matches the
large comoving correlation length typical of evolved EROs at
$z\simeq$1 and of clusters of galaxies at $z=0$ (Figure
\ref{fig:blain}). From this figure, we see that the correlation
function of SMGs appears to be consistent with the hypothesis that
they are associated with the most massive dark matter halos at high
redshifts. These are more massive than the host halos of LBGs and QSOs
at comparable redshift.  These preliminary conclusions have to be
investigated in more detail.  In particular, the SMG masses inferred
from clustering measurements have to be compared to the dynamical
masses derived from millimeter wave CO spectroscopy.  To go deeper in
understanding the clustering properties of infrared galaxies, {\it
Spitzer} 24~$\mu$m surveys will be best suited. With a large number of
sources detected in quite large surveys, {\it Spitzer} will
unambiguously constrain the clustering of infrared galaxies from
redshift 0.5 to 2.5. Conjointly, the physics of galaxy clustering can
be probed by the CIB fluctuation analysis (Knox et al. 2001). CIB
fluctuations measure, on large angular scales, the linear clustering
bias in dark-matter halo and, at small angular scales, the nonlinear
clustering within a dark-matter halo (Cooray \& Sheth 2002).  They
thus probe both the dark-matter halo mass scale and the physics
governing the formation of infrared galaxies within a halo. Promising
attempts are underway using the 170~$\mu$m ISO FIRBACK fields (Lagache
et al., in preparation).

\begin{figure*}[!ht]
\begin{center}
\includegraphics[width=1.0\linewidth]{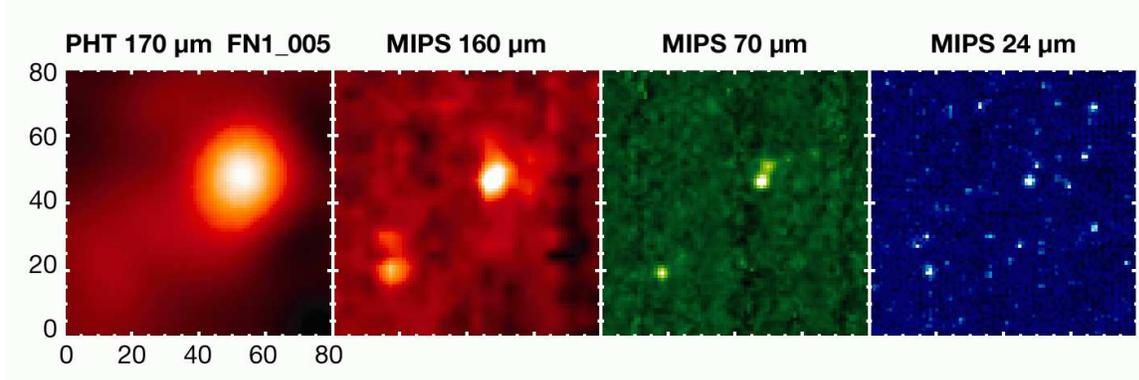}
\end{center}
\caption{Effects of confusion in the far-infrared. Observation of a
source in the ELAIS-N1/FIRBACK field in a 400x400 square arcsec
box. All images have been resampled to 5 arcsec per pixel, which
oversamples the far-infrared maps but undersamples the mid-infrared map.
From left to right:
373~mJy ISOPHOT 170~$\mu$m source with about 128s of integration (FIRBACK
survey, Dole et al. 2001); labels indicate the 5
arcsec pixels;
{\it Spitzer}/MIPS 160~$\mu$m with about 16s of integration
(SWIRE survey, Lonsdale et al. 2004);
MIPS 70~$\mu$m with about 80s of integration (SWIRE);
MIPS 24~$\mu$m with about 160s of integration (SWIRE).
Notice (1) the ISO 170~$\mu m$ source is marginally resolved
with MIPS 160, and is unambiguously resolved at 70~$\mu$m and 24~$\mu$m;
(2) the two fainter MIPS 160~$\mu$m resolved sources (bottom left) create
fluctuations in the ISO 170~$\mu$m map that produce the confusion
noise when the resolution is limited.
}
\label{fig:confusion}
\end{figure*}

\section{\label{highz} FINDING ULIRGs AT $z>$3: CONFUSION}
 
In the near future, when a proper census of ULIRGs up to $z\simeq$3
will have been carried out, the fraction of the CIB at $\sim$1~mm not accounted for should
give an indication of the contribution from sources at larger
redshifts.  Deep surveys at $\sim$1-2~mm are the only obvious tool to find
most of these sources. However, the limiting factor of the surveys is not only
detector sensitivity or photon noise but also confusion. 
We concentrate in this section only on extragalactic sources confusion.
Source confusion in the far-infrared is illustrated in Figure \ref{fig:confusion}.\\

\begin{figure*}[!ht]
\begin{center}
\includegraphics[width=1.0\linewidth]{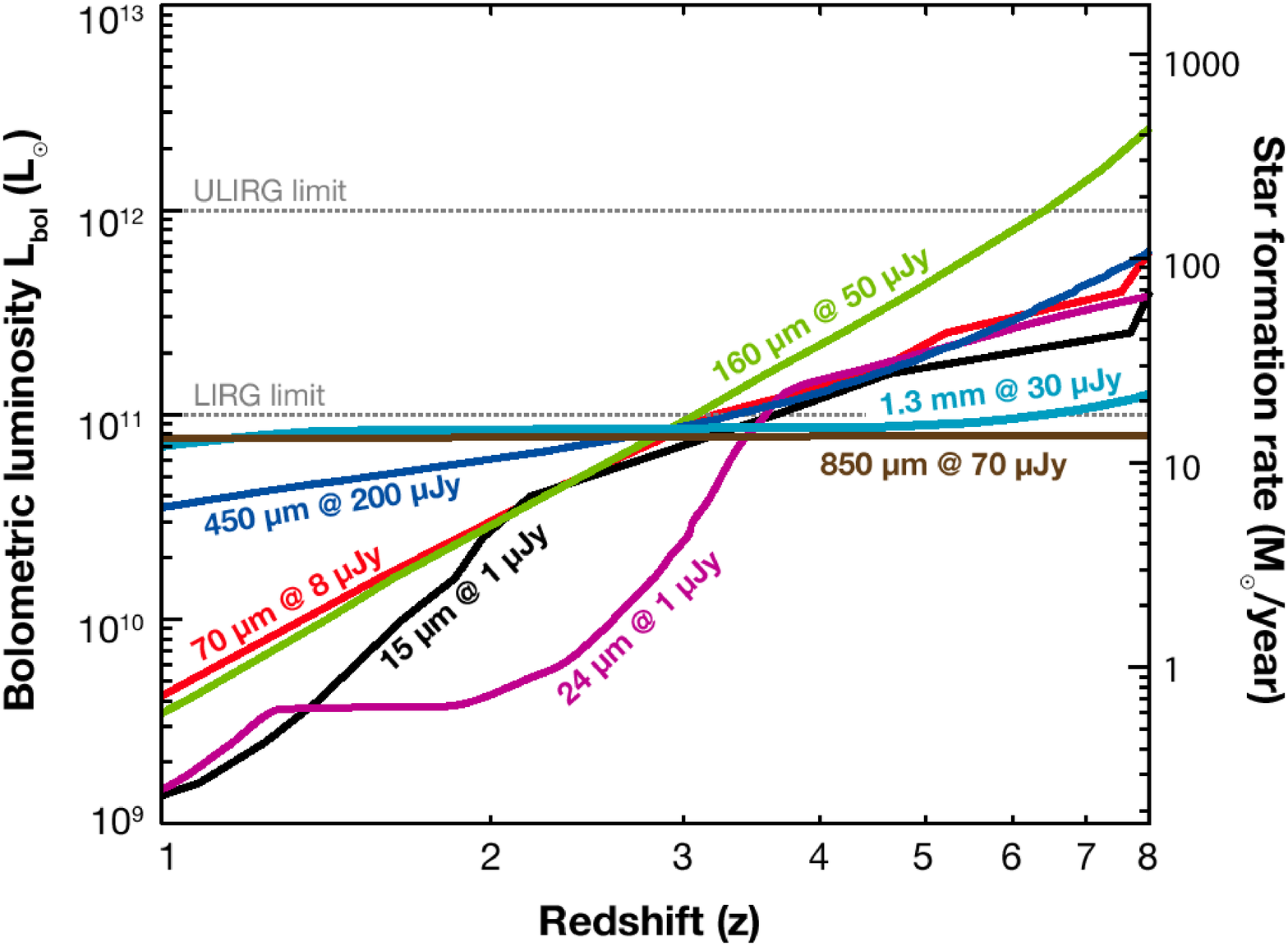}
\end{center}
\caption{
Sensitivity to the bolometric luminosity (and star-formation
rate, assuming star-forming galaxies) of hypothetical surveys
designed to detect LIRGs at $z \sim 3$. The required sensitivities are:
at 15~$\mu$m: $S_{\nu} > 1 \mu$Jy;
at 24~$\mu$m: $S_{\nu} > 1 \mu$Jy;
at 70~$\mu$m: $S_{\nu} > 8 \mu$Jy;
at 160~$\mu$m: $S_{\nu} > 50 \mu$Jy;
at 450~$\mu$m: $S_{\nu} > 200 \mu$Jy;
at 850~$\mu$m: $S_{\nu} > 70 \mu$Jy;
at 1.3 mm: $S_{\nu} > 30 \mu$Jy.
This plot makes use of the Lagache et al. (2004) model (see appendix).
}
\label{fig:plot_lbol_vs_z4araa_future}
\end{figure*}

Predicting or measuring confusion depends on the scientific goal of
the measurement (Helou \& Beichman 1990; Dole et al. 2003; Lagache et
al. 2003). Performing an unbiased far-infrared or submillimeter survey
and getting a complete sample has different requirements than
following-up in the far-infrared an already known near-infrared source
to get an SED and/or a photometric redshift. In the former case, one
has to tightly control the statistical properties of the whole sample;
in the latter case, completeness is irrelevant, and even a low
photometric accuracy is adequate. We thus favor the use of a term like
``unbiased confusion'' for the former case. New techniques are being
developed to use a priori information at shorter wavelength (e.g.,
8~$\mu$m with {\it Spitzer}/IRAC and 24~$\mu$m with MIPS) to infer
some statistical properties (such as source density or SED) of sources
at longer wavelength (e.g., 24 or 160~$\mu$m, respectively), and thus
to beat unbiased confusion.  Predicting unbiased confusion (for
instance Condon 1974; Franceschini et al. 1989; Helou \& Beichman
1990; Dole et al. 2003; Lagache et al. 2003; Takeuchi et al. 2004;
Negrello et al. 2004) requires the knowledge of at least the
number-count distribution of the galaxies. In practice, models
(validated at some point by observations) are used. Because the shape
of the counts in a $log(N) - log(S_{\nu})$ diagram varies with the
flux density $S_{\nu}$, the fluctuation level of faint sources below
$S_{\nu}$ also vary.  This fluctuation level gives an estimate of the
unbiased confusion using a photometric criterion (Lagache \& Puget
2000; Dole et al. 2003; Lagache et al.  2003). At very faint fluxes,
when the background is almost resolved, the photometric criterion will
obviously give a very small value for the unbiased confusion level,
but the observations will be limited by the confusion due to the high
density of faint resolved sources. Thus, another criterion, the source
density criterion for unbiased confusion (SDC, Dole et al. 2003; Dole
et al. 2004b), needs to be computed and compared to the photometric
criterion.

In the infrared and submillimeter range below 300~$\mu$m, the unbiased
confusion is in general better predicted by the source density
criterion for current and future facilities, because the angular
resolution has improved (e.g. from ISO to {\it Spitzer}). At longer
wavelengths, the photometric criterion is more useful.  We use the
model of Lagache et al. (2004) to predict unbiased
confusion. Fifteen-meter submillimeter telescopes are limited by the
confusion at 1.2~mJy at 850~$\mu$m and at 0.5~mJy at 1.2~mm.  Ongoing
surveys (Smail et al. 2002; Greve et al. 2004) already reach or are
about to reach these levels. If we want to resolve about 80\% of the
CIB, which corresponds to 56~$\mu$Jy at~850 $\mu$m and 20~$\mu$Jy at
1.3~mm, one would need a $\sim$90-m telescope at 850~$\mu$m and a
$\sim$150-m telescope at 1.4~mm.  Future facilities for infrared and
submillimeter observations include far-infrared space observatories
such as {\it Herschel}, SPICA, and SAFIR, survey missions like ASTRO-F
and {\it Planck}, a larger near-infrared and mid-infrared observatory,
{\it JWST}, and a ground-based submillimeter interferometer, ALMA.  In
order to detect LIRGs at $z \sim 3$, experiments operating at 15, 24,
70, 160, 450, 850, 1380~$\mu$m respectively should reach a sensitivity
of 1, 1, 8, 50, 200, 70, 30~$\mu$Jy, respectively (see Figure
\ref{fig:plot_lbol_vs_z4araa_future}). The spectral window around
450~$\mu$m seems the most effective to reach these galaxies.  This
constraint is somewhat relaxed if one wants to detect ULIRGs at $z\sim
3$, in which case the required sensitivities are multiplied by about
10.\\ In the near future, ASTRO-F, {\it Herschel}/SPIRE and {\it
Planck} will be mostly limited by confusion. At long wavelengths, to
probe most of the CIB source population and to detect enough early
mergers made by building blocks not yet affected by star formation and
evolution, large extragalactic surveys will have to be conducted with
ALMA. These surveys will take a substantial fraction of the time
(Lagache et al. 2003). As an example, mapping one square degree at
1.3~mm at the 5$\sigma$ sensitivity of 1~mJy -- $\sim$50\% of the CIB
is resolved -- takes 138 days without including overheads.

\section{COMPARISON OF MODELS WITH OBSERVATIONS}

One of the striking result of the deep surveys concerns the evolution
of the infrared and submillimeter galaxy population.  The source
counts are high when compared to no evolution, or moderate, evolution
models{\footnote{`No-evolution': the co-moving luminosity function
remains equal to the local one at all redshifts}} for infrared
galaxies.  Classical semianalytical models of galaxy formation
predicts neither the large numbers of infrared galaxies nor their very
strong evolution, revealing a serious gap in our understanding of
galaxy formation and evolution.  Very recently, several empirical
approaches have been proposed to model the high evolution of the
infrared output with redshift (e.g., Chary \& Elbaz 2001; Franceschini
et al. 2001; Rowan-Robinson 2001; Takeuchi et al. 2001; Xu et
al. 2001; Lagache et al. 2004) that fit source counts, redshift
distributions and CIB intensities and fluctuations, although often not
all of them. All these models, however, agree on a general trend --
i.e., the luminosity function must change dramatically with redshift,
with a rapid evolution of the high-luminosity sources (L$>$2 10$^{11}$
L$_{\odot})$ from $z=0$ to $z=1$, which then stay rather constant up
to redshift 3 or more.  The evolution of the infrared luminosity
function may be linked to a bimodal star-formation process, one
associated with the quiescent and passive phase of the galaxy
evolution and one associated with the starburst phase, triggered by
merging and interactions.  The latter dominates the infrared and
submillimeter energy density of the Universe at high $z$.
Consistently, cold dark matter N-body simulations show that halo
merger rates increase with redshift as (1+z)$^m$ with 2.5$\le$ m
$\le$3.5 (Gottlober et al. 2001).  Observations, however, give $m$
values between 0 and 4 (Le F\`evre et al. 2000; Conselice et al. 2003;
Bundy et al. 2004; Lin et al. 2004). The spread is due to different
selection effects, detection techniques, pair criteria and sample
variance. It is therefore not easy to reconcile the different
observational results. Moreover, comparisons with models are very
difficult because definitions of merger rates may not be
consistent. Merger rates can also depend on halo masses.  As a
consequence, the timescale of the merger phase is difficult to
estimate. Peaks of star formation produced by mergers in
hydrodynamical models (e.g., Scannapieco \& Tissera 2003) has a
duration of several hundreds of million years. This is consistent with
what is observed.  ULIRGs emit more than half of their bolometric
luminosity from a starburst of age 10$^7$-10$^8$ years (Genzel et
al. 1998). LIRGs build up their stellar mass in a typical timescale of
about 0.1 Gyr (Franceschini et al. 2003). These timescales are also
supported by Marcillac et al. (2005) who performed Monte Carlo
simulations using synthetic spectra based on the models of Bruzual and
Charlot (2004) to derive the past star-formation history of 22
LIRGs. They found that LIRGs experience a major event of star
formation in their lifetime that produce about 10\% of their stellar
mass within 0.1 Gyr. How many such episodes of violent star formation
does a typical galaxy experience? Assuming a timescale of 0.1 Gyr,
Hammer et al. (2005) estimate the number of episodes per galaxies as
about 5 from $z=1$ to $z=0.4$. These episodic bursts naturally explain
the high fraction of LIRGs in the distant Universe.\\

Models that are more sophisticated than empirical approaches attempt
to follow the physics of galaxy formation in greater detail (e.g.,
Guiderdoni et al. 1998; Hatton et al. 2003; Granato et al. 2004; Silva
et al. 2005).  In semianalytical models, the collapse of perturbations
is described by the classical top-hat model under the assumptions of
homogeneity and sphericity. The mass distribution of collapsed halos
is computed from the so-called peaks formalism developed by Bardeen et
al. (1986). Then dissipative collapse and cooling are introduced, with
the usual ``overcooling'' problem that can partly be solved by
introducing stellar feedback. Star-formation processes are deduced
from the gas content and the dynamical timescale of the galaxies.
Finally spectrophotometric evolution is used to compute the age
dependence of the gas content, the spectra of the stellar populations
and the mass-to-luminosity ratios. To make specific predictions for
the infrared galaxies, these models must include an important
additional feature: absorption of the UV/optical radiation and
emission by the dust grains.  Very often two modes of star formations
are considered; a quiescent mode and a burst mode in which the star
formation timescales are much shorter.  This burst mode is triggered
by galaxy mergers and is absolutely required by the infrared to
submillimeter observations. There are some indications that to
reproduce the submillimeter galaxy counts, a dramatic change of the
IMF is required. A top-heavy IMF, in particular, increases the
production of dust that is essential for boosting the luminosity of
galaxies in the submillimeter. Using an IMF of the form $dN$/dln$m
\propto m^{-x}$ with $x$=0 for the burst mode, Baugh et al. (2005)
were able to reproduce not only the submillimeter observations but
also the properties of Lyman-Break galaxies. They predict that the
SMGs reside in the more massive halos in place at $z=2$ and therefore
that they are more strongly clustered than dark matter at this
epoch. This is consistent with tentative observational constraints
(Blain et al. 2004a). There are several observational ``indications''
of massive stars ($>$ 100 M$_{\odot}$) in nearby starburst templates.
Wolf-Rayet stars\footnote{Wolf-Rayet stars are hot (25,000 to 50,000
K), massive ($\ge 25 M_{\odot}$), luminous stars with a high rate of
mass loss. The Wolf-Rayet phase appears in an advanced stage of
evolution. They are believed to be O stars that have lost their
hydrogen envelopes, leading their helium cores exposed. Wolf-Rayet
stars are often in a binary system, and are deemed, within a few
million years, to explode as type Ib or Ia supernovae.} have been
detected in a large number of galaxies undergoing intense bursts of
star formation (e.g., Gonzalez-Delgado et al. 1997; Pindao et
al. 2002). However, it remains difficult to measure the IMF at high
mass because of aging effects that can mimic real upper-mass IMF
cutoff (the highest massive stars have very short lifetimes).
\\
In conclusion, the hierarchical galaxy formation paradigm is very
successful in its description of large-scale structure formation and
evolution.  The next important step will be to test this picture to
explain not only the number densities but also the mass assembly and
particularly the mass of the SMGs.  First mass measurements of SMGs
galaxies seem to show that a very flat IMF cannot by itself explain
the mass assembly of the baryonic matter at high $z$ (Genzel et
al. 2004). Hierarchical clustering underpredicts the high-$z$ volume
densities of these massive galaxies. More work needs to be done to
test the baryonic mass assembly in the hierarchical paradigm. Both
observational and model estimates are still very uncertain, with the
former depending on large lifetime corrections and small samples and
the latter on ad hoc input recipes for feedback and star formation.

\section{CONCLUSION AND OPEN QUESTIONS}

A number of conclusions are now clear from the analysis of the identified 
sources in the CIB:
\begin{itemize}

\item The comoving energy produced in the past that makes up the CIB at different wavelengths 
is more uniform that what is suggested by its spectral energy
distribution. This is due to the fact that the CIB at long wavelengths
($\lambda \ge 400 \mu$m) is dominated by emission from the peak of the
SED of galaxies at high $z$. More quantitatively, the ISOCAM surveys
reveal that about two-thirds of the CIB emission at $\lambda \sim$150
$\mu$m is generated by LIRGs at $z \sim 0.7$. At 850~$\mu$m, more than
half of the submillimeter CIB is generated by SMGs. The brightest SMGs
($S_{850}>$3 mJy, $\sim$30\% of the CIB) are ULIRGs at a median
redshift of 2.2. The energy density at 150~$\mu$m, which is
$\sim$20-25 times larger than the energy density at 850~$\mu$m
requires a comoving energy production rate at $z=0.7$ roughly 10 times
the energy production rate at $z=2.2$.

\item The evolution exhibited by LIRGs and ULIRGs is much faster than for optically 
selected galaxies. The ratio of infrared to optical, volume-averaged
output of galaxies increases rapidly with increasing redshift.

\item Luminosity function evolution is such that the power output is dominated by
LIRGs at $z\simeq 0.7$ and ULIRGs at $z\simeq 2.5$.

\item The energy output of CIB sources is dominated by starburst activity.

\item AGN activity is very common in the most luminous of these galaxies even though this 
activity does not dominate the energy output. The rate and fraction of
the energy produced increase with the luminosity.

\item LIRGs at $z\simeq 0.7$ are dominated by interacting massive
late-type galaxies. Major mergers become dominant in ULIRGs at
$z\simeq 2.5$.

\item SMGs show rather strong correlations with correlation lengths larger than those of other
high redshift sources.

\item LIRGs and ULIRGs cannot be identified with any of the distant 
populations found by rest-frame ultraviolet and optical surveys.

\end{itemize}

Although these findings are answering the basic questions about the
sources that make up the CIB, there are still observational
difficulties to be overcome to complete these answers.  The SEDs of
LIRGs and ULIRGs are quite variable and often not very well
constrained in their ratio of far-infrared to mid-infrared or to
submillimeter wavelengths.  The far-infrared, where most of the energy
is radiated, requires cryogenically cooled telescopes. These have
small diameters and, hence, poor angular resolution and severe
confusion limits for blind surveys. Establishing proper SEDs for the
different classes of infrared galaxies detected either in mid-infrared
(with ISOCAM at 15~$\mu$m or MIPS at 24~$\mu$m) or in
millimeter-submillimeter surveys is one of the challenges of the
coming decade.  Making sure that no class of sources that contribute
significantly to the CIB at any wavelength has been missed is an other
observational challenge. The submillimeter galaxies not found through
the radio-selected sources and the question of the warm submillimeter
galaxies are also two of those challenges.

Multiwavelength observations of high-$z$ infrared galaxies give a
number of new insights on the galaxy formation and evolution
problem. As an example, the gas masses and total masses of SMGs are
found to be very high. There is a first indication that the number of
such high-mass object at redshifts between 2 and 3 is uncomfortably
large compared to semianalytical models of galaxy formation based on
the standard hierarchical structure-formation frame. The evolution of
the luminosity function is dominated by more luminous sources as
redshift increases. This is surprising because the mass function of
the collapsed structure is expected to be dominated by smaller and
smaller objects as redshift increases.

The populations of infrared galaxies concentrated at $z\simeq 1$ and
at $z\simeq 2.5$ studied so far reveal rather different type of
sources.  The lower redshift ones seem to be starburst phases of
already-built massive, late-type field galaxies accreting gas or
gas-rich companions forming the disks. We see today a rapid decrease
of this activity probably associated with a dry out of the gas
reservoir in their vicinity.  The larger redshift ones, which are also
more luminous, seem to belong to more massive complex systems
involving major merging. These systems could be located in the rare
larger amplitude peaks of the large-scale structures leading to
massive elliptical galaxies at the center of rich clusters. The
redshift distribution of these seems quite similar to the redshift
distribution of quasars.

Interesting problems that are central to the understanding of galaxy formation and evolution have
to be solved in the next decade:
\begin{itemize}
\item  Determining the role of the large scale environment (nodes, filaments and sheets 
of the large-scale structures) on star formation; 
\item Find the relative rates of accretion of gas and smaller galaxies
in the growth of massive objects;
\item Establish the cycle of bulge versus disk formation, as a function of the ratio 
between stars and gas in the accreted material;
\item Identify the different types of starburst (in a disk or in the nucleus, interaction or merger 
driven); and
\item Estimate the fraction of time spent in the starburst phase and the duration of this phase
\end{itemize}
Current observations all point in the direction of a 
possible strong effect of the large-scale environment and the need for models of hierarchical formation 
and evolution that include properly star versus gas ratio in the accreted material. 

Finally the connection between the starburst phenomenon and the AGN activity is an old question still largely
unresolved. Recent observations of infrared/submillimeter galaxies have
reinforced the link but have not much improved our understanding of the physical link.
 
\par\bigskip\noindent
\par\bigskip\noindent
{\bf ACKNOWLEDGMENTS}
\par\smallskip\noindent
We are very grateful to Alexandre Beelen, Karina Caputi, David Elbaz,
Fran\c{c}ois Hammer, and George Helou for very useful discussions
during the writing of this manuscript. We also thank Pierre Chanial
for providing us the M82 SED.  Finally, we warmly thank the scientific
editor, who found a large number of typos and mistakes in our use of
English. The reading of the paper has been significantly improved by
his detailed corrections.

\par\bigskip\noindent
\par\bigskip\noindent
{\bf APPENDIX: THE PHENOMENOLOGICAL MODEL OF LAGACHE ET AL. (2004)}
\par\smallskip\noindent
In this paper, we make extensive use of the Lagache et al. (2004)
phenomenological model to illustrate our points.  This model
constrains in a simple way the evolution of the infrared luminosity
function with redshift.  It fits all the existing source counts
consistent with the redshift distribution, the CIB intensity, and, for
the first time, the CIB fluctuation observations, from the
mid-infrared to the submillimeter range.  In this model, Lagache et
al. (2004) assume that infrared galaxies are mostly powered by star
formation and hence they use SEDs typical of star-forming galaxies.
Although some of the galaxies will have AGN-dominated SEDs, they are a
small enough fraction that they do not affect the results
significantly.  They therefore construct ``normal'' and starburst
galaxy template SEDs: a single form of SED is associated with each
activity type and luminosity.  They assume that the luminosity
function is represented by these two activity types and that they
evolve independently. They search for the form of evolution that best
reproduces the existing data.  An example of two cosmological
implications of this model is $(a)$ the PAH features remain prominent
in the redshift band 0.5--2.5 (as observationally shown by e.g.,
Caputi et al. 2005), and $(b)$ the infrared energy output has to be
dominated by $\sim$3$\sim$10$^{\rm 11}$~L$_{\rm \odot}$ to
$\sim$3$\times$10$^{\rm 12}$~L$_{\rm \odot}$ galaxies from redshift
0.5 to 2.5.\\ The excellent agreement between the model and all the
available observational constraints makes this model a likely good
representation of the average luminosity function as a function of
redshift and a useful tool to discuss observations and models. Its
rather simple assumptions such as the single parameter sequence of
SEDs for starburst galaxies is certainly not accounting for some of
the detailed recent observations but probably do not affect seriously
the redshift evolution of the averaged properties which are what is
modeled.


\end{document}